# Spin-dependent magnetism and superparamagnetic contribution to the magnetocaloric effect of non-stoichiometric manganite nanoparticles

Nikita A. Liedienov [a, b], Ziyu Wei [a], Viktor M. Kalita [c, d, e], Aleksey V. Pashchenko [a, b, d, *], Quanjun Li [a], Igor V. Fesych [f], Vitaliy A. Turchenko [g, h], Changmin Hou [i], Aleksey T. Kozakov [j], Georgiy G. Levchenko [a, b, *]

[a] *State Key Laboratory of Superhard Materials, International Center of Future Science, Jilin University, 130012 Changchun, P. R. China*
[b] *Donetsk Institute for Physics and Engineering named after O.O. Galkin, NAS of Ukraine, 03028 Kyiv, Ukraine*
[c] *National Technical University of Ukraine "Igor Sikorsky Kyiv Polytechnic Institute", 03056 Kyiv, Ukraine*
[d] *Institute of Magnetism, NAS of Ukraine and MES of Ukraine, 03142 Kyiv, Ukraine*
[e] *Institute of Physics, NAS of Ukraine, 03028 Kyiv, Ukraine*
[f] *Taras Shevchenko National University of Kyiv, 01030 Kyiv, Ukraine*
[g] *Frank Laboratory of Neutron Physics, Joint Institute for Nuclear Research, 141980 Dubna, Russia*
[h] *South Ural State University, 454080 Chelyabinsk, Russia*
[i] *State Key Laboratory of Inorganic Synthesis and Preparative Chemistry, College of Chemistry, Jilin University, 130012 Changchun, P. R. China*
[j] *Scientific-Research Institute of Physics at Southern Federal University, 344194 Rostov-na-Donu, Russia*

These authors contributed equally: N.A. Liedienov, Ziyu Wei
* Corresponding author
*E-mail addresses*:  alpash@ukr.net (Aleksey V. Pashchenko)
   g-levch@ukr.net (Georgiy G. Levchenko)

## Abstract

Despite extensive researches on manganites owing to widespread use in modern electronics, this class of metal oxides does not cease to surprise with its unique properties and new phenomena. Here we have studied structural and magnetic properties of non-Heisenberg manganite nanoparticles with a strong spin-electron coupling. During transition from low temperature ferromagnetic to high temperature paramagnetic state, the change in charge, valence and spin (magnetic moment) with the localization of $e_g$-electrons on the manganese ions have been detected. With decrease in a temperature, the overstated effective magnetic moment of Mn ions in paramagnetic phase is reduced dramatically testifying to spin-dependent magnetism. The critical behavior of magnetization with determination of critical parameters near second-order phase transition has been studied comprehensively. Based on unusual behavior of temperature and field dependences of magnetic entropy change under different magnetic field, an additional influence of superparamagnetism of nanoparticles on the magnetocaloric effect has been found.

# 1. Introduction

Since discovery of magnetocaloric effect (MCE), the search of new materials with a large magnetic entropy change $-\Delta S_M$ has become topical and priority owing to their potential applications as green and environmentally friendly magnetic refrigerators [1-3]. $Gd_5(Si_2Ge_2)$, $La(Fe_{1-x}Si_x)$ and $MnAs_{1-x}Sb_x$ intermetallic compounds, Ni–Mn–X (X = Ga, In, Sn) Heusler alloys, Gd-based alloys, $ErCo_2$, $Ho_2O_3$, *etc.* demonstrate excellent MCE values [4-11]. However, most of these materials have a number of disadvantages, such as: expensive elements, complex synthesis route, a large thermal and magnetic hysteresis, *etc.* [12]. Other potential candidates devoid of these shortcomings with low production cost, high chemical stability, simple preparation way and low eddy current heating are metal oxides based on the manganites with the large MCE values [13].

Manganites are good candidates for MCE. They can show a significant magnetic entropy change in low magnetic fields near room temperature.

Magnetic manganese ions in manganites can have different valences and, as a consequence, different spin values (quantum number) characterizing their ionic states. In addition, the charge and spin states of ions are sensitive to defect chemistry. Therefore, manganites are not classical magnets, with traditional pair exchange between neighboring ions, where the spin of the ions is constant regardless of the temperature and type of the magnetic phase. In this work, we show that a change in the spin state of Mn ions can be observed in manganites during a change in the type of magnetic order upon heating or cooling of the samples, which occurs without changing the crystal structure of the sample. This is the first time we pay attention to this and note that it is important for understanding the magnetism of these magnetic materials.

The general formula of manganites with a perovskite $ABO_3$ structure is $A_{1-x}A'_xMnO_3$, where $A$ = rare-earth elements and $A'$ = alkaline or alkaline rare-earth elements [14]. During non-isovalent replacing $A$ by $A'$ ions, the mixed-valence state of manganese ($Mn^{3+}/Mn^{4+}$) can be obtained with different strength of the exchange interactions, such as: ferromagnetic (FM) double-exchange (DE) interaction *via* the $Mn^{3+} \leftrightarrow O^{2-} \leftrightarrow Mn^{4+}$ and antiferromagnetic (AFM) super-exchange interaction *via* the $Mn^{3+} \leftrightarrow O^{2-} \leftrightarrow Mn^{3+}$ and $Mn^{4+} \leftrightarrow O^{2-} \leftrightarrow Mn^{4+}$ [15]. Introducing monovalent *A'* elements leads to the biggest change in the $Mn^{3+}/Mn^{4+}$ ratio and, consequently, the finest tune of magnetic phase transition temperature and magnetocaloric parameters [12]. Availability of anion vacancies $V^{(a)}$ in oxygen sublattice $O_{3-\Delta}$ weakens DE, but the magnetic entropy change may persist [16]. Moreover, the creation of cation vacancies $V^{(c)}$ in *A*-sublattice influences significantly the magneto-transport properties of manganites modifying bond lengths Mn–O and angles Mn–O–Mn, which are responsible for the strength of exchange interactions, as well as leads to appearance of $Mn^{2+}$ ions on *A*-sites and an additional multiple FM DE interaction *via* $Mn^{3+} \leftrightarrow O^{2-} \leftrightarrow Mn^{2+} \leftrightarrow O^{2-} \leftrightarrow Mn^{4+}$ [17-21]. Depending on *A* or *A'* deficient sublattices, the Curie temperature $T_C$ and magnetocaloric parameters can be tuned precisely [18, 22-24]. In addition, a change in content of Mn on *A*- and *B*-sites by the substitution for other ions, creation cation vacancies and/or introduction of overstoichiometric manganese also strongly influences the magnetotransport properties of manganites [18-21, 25, 26]. The replacement of Mn on *B*-sites by other 3*d*-elements usually leads to decrease in the $T_C$ and magnetocaloric parameters due to weakening the DE interaction and appearance of additional FM and/or AFM interactions between $3d\text{-ion} \leftrightarrow O^{2-} \leftrightarrow Mn^{3+/4+}$ [27-29]. Introducing excess Mn on *A*- and/or *B*-positions has a number of advantages since it brings to the completeness of *B*-sublattice due to existence of $V_B^{(c)}$ cation vacancies and significantly improves the magnetoresistance effect without lowering the $T_C$ [30]; increases the metal-insulator temperature [19]; enhances the transport properties and FM metallic state due to appearance of $Mn^{2+}$ ions on *A*-positions, which have half-filled conduction band crossing the Fermi level and cause multiple DE [20]; increases the $T_C$ and MCE [12, 31-33].

Despite a lot of publications on the study of the magnetic and magnetocaloric properties of manganites with above mentioned variations in composition, there are open and unsolved questions associated with definition of the charge and spin states of magnetic ions. In this paper, the variation in the spin state of ions during a change in the magnetic order is observed in manganites that is an illustration of a non-Heisenberg-type magnet with atypical critical behavior near Curie temperature.

Nowadays, there are plenty of theories, such as: DE interaction [34], Griffiths phase [35], electron-phonon coupling [36], critical phenomena [32, 37-39], which do not take into account the change in the magnitude of the spin of Mn ions during decreasing or increasing temperature without change in a structure. We will experimentally show the existence of this effect in manganites.

The increased interest to manganites is also associated with their special transport properties due to the strong spin-electron correlations [14, 40], which is another reason for the strong nonlinearity of the magnetic properties of manganites associated with the charge state of their magnetic ions. The phase transition from the metal type of conductivity to the isolator state is associated with the localization of $e_g$-electrons on the Mn ions [41] that should lead to a change of the spin value in the high temperature region.

Thus, the change in the charge and spin state of Mn ions during a change in the magnetic order caused by strong spin-electron interactions means that the phenomenon of spin-dependent magnetism should occur in manganites during the transition from paramagnetic (PM) to FM state. The spin quantum number, $S$, of Mn ions in PM and FM phase can be different, which will be obtained in this paper.

It should be noted that traditionally in magnets it is always assumed that the quantum number (spin) is a constant, $S$ = const, and during magnetic ordering, only the magnitude of the spin's projection changes. In manganites, it is turned out that the spin of ions is different in PM and FM phases, $S(T) \neq$ const, without any structural phase transitions, and only the magnetic symmetry changes.

The spin-dependent magnetism of manganites can also be one of the reasons for enhancing the non-Heisenberg character of interionic spin-spin interactions, and, as a consequence, lead to an increase in the MCE in low magnetic field. The non-Heisenberg addition to the exchange field enhances the effect of the external magnetic field and leads to a stronger change in the magnetic entropy upon magnetization. This should lead to an additional contribution to the MCE from fluctuations of the SPM type generated by the thermal disalignment of the magnetic moments of nanoparticles with its contribution to the magnetic entropy, which, as it turned out in the studied samples, is not small and has a temperature feature near the Curie point.

Thus, in this work, we have synthesized nanopowder of non-stoichiometric $La_{0.8-x}\square_{x}Na_{0.2}Mn_{1+x}O_{3-\Delta}$ manganite which simultaneously depends on three chemical parameters: (i) concentration of cation $V_A^{(c)}$ vacancies "$\square$"; (ii) concentration of anion $V^{(a)}$ vacancies "$O_{3-\Delta}$"; and (iii) concentration of overstoichiometric manganese "$Mn_{1+x}$". The influence of these parameters on the structure, morphology, valence, charge and spin state, as well as magnetic and magnetocaloric properties of the $La_{0.8-x}\square_{x}Na_{0.2}Mn_{1+x}O_{3-\Delta}$ compounds has been studied. The change in spin value of Mn ions depending on the both temperature region and their magnetic state as well as an additional SPM contribution of nanoparticles to the MCE have been discovered.

## 2. Experimental

### 2.1. Sample preparation

Nanopowders of non-stoichiometric $La_{0.8-x}\square_{x}Na_{0.2}Mn_{1+x}O_{3-\Delta}$ manganite with a concentration of $x = 0.00, 0.05, 0.10, 0.15$ and $0.20$ were synthesized using a sol-gel method [32, 42] from the initial $La(NO_3)_3 \cdot 6H_2O$ (purity 99.9%), $NaNO_3$ (purity 99.0%) and $Mn(NO_3)_2 \cdot 4H_2O$ (purity 98.0%) components. The calculated amount of nitrates and citric acid (the ratio of the molar sum of metals and $H_3Cit \cdot H_2O$ was 1:1) were dissolved in deionized water. The solution was evaporated up to formation of a homogeneous gel, which was dehydrated and heated gradually from 200 to 500 °C (the rate of heating was 50 °C/h). The retention time at a temperature of 500 °C was 5 hours. The resulting black powder was ground in an agate mortar using i-PrOH, then placed in platinum crucibles and synthesized at 850 °C (20 h). The samples were cooled in the regime of natural heat exchange between furnace and environment in air.

*2.2. Characterization*

Oxygen content "$O_{3-\Delta}$" in all compositions was determined using IT method [43, 44]. The $La_{0.8-x}\square_{x}Na_{0.2}Mn_{1+x}O_{3-\Delta}$ nanopowders ($m$ = 50 mg) were placed in flask adding 10 mL of HCl solution (concentration of 0.7 mol / L) and 10 mL of KI solution (concentration of 1 mol / L). The dissolution of the samples was about of 10–30 minutes. The iodine formed after dissolution of the substituted manganite powder was titrated with a $Na_2S_2O_3$ solution (concentration of 0.1 mol / L) using a freshly prepared starch solution as an indicator. Accuracy in determination of the oxygen content was ±0.02 per formula unit.

The phase composition, lattice parameters, and size of coherent scattering regions were defined by X-ray diffraction method using Shimadzu LabX XRD-6000 diffractometer in $CuK_{\alpha 1}$-radiation ($\lambda$ = 0.15418 nm) at room temperature. The structural refinement was performed with Rietveld analysis using the FullProf software. An average size of coherent scattering regions $D$ was determined using the Debye-Scherrer method with an accuracy ± 1nm [45].

The morphology and size of the particles were determined using JEM-2200FS Transmission Electron Microscope. High-resolution transmission electron microscopy with accelerating voltage 200 kV was employed to obtain information about size and shape of the particles, as well as to determine an average interplanar distance using Gatan Microscopy Suite software based on the fast Fourier transformation. The samples for the TEM analysis were prepared by placing a drop of diluted mixture of particles and acetone on carbon coated copper grid. An average particle size $d$ was obtained from analysis of TEM images during approximation of the experimental values of $d$ by different distribution functions. The values of particle size $d$ were measured within clear and defined particle margins using Nano Measure 1.2.5 software [42]. It should also be noted that obtaining clear TEM images was quite complex because of magnetic attraction between the nanoparticles and their agglomeration since the manganite nanopowder has the Curie temperature which is higher than room temperature (see S4). Moreover, the chemical composition of all samples has performed by EDS using

Bruker XFlash 6|60 in the Hitachi Regulus 8100 Scanning Electron Microscope analyzing at least 10 areas of around 200 $\mu$m for every concentration.

X-ray photoelectron spectra of the $La_{0.8-x}\square_{x}Na_{0.2}Mn_{1+x}O_{3-\Delta}$ nanopowders were obtained at room temperature using XPS with an ESCALAB 250 X-ray photoelectron microprobe. The spectra were excited with monochromatized radiation of $AlK_{\alpha}$. The state of the surfaces was monitored with a C1s line. Its intensity was very small, but the line was still detectable in the background, which allowed calibrating the energy scales for all of the spectra. The C1s line binding energy was about 285 eV. Fine layers of the $La_{0.8-x}\square_{x}Na_{0.2}Mn_{1+x}O_{3-\Delta}$ nanopowders were deposited on double-sided conductive adhesive tape. The flow of slow electrons was used to counteract the charging of the powders. The background of the X-ray photoelectronic lines was cut off using the Shirley method [46]. It should also be noted that the possibility of the XPS beam penetration $\sim 3\lambda$ [46], where the $\lambda$ is the free path length of electrons depending on the energy of the electronic line that, for example, for Mn ions is $\sim 5.8$ nm.

The magnetic measurements were performed using Quantum Design SQUID MPMS 3 in a temperature range of $T = 2$–400 K and magnetic field $H$ up to 3 T. The magnetic properties were determined based on the analysis of the temperature $M(T)$ and field $M(H)$ dependencies of the magnetization.

The critical temperature and exponents were determined using Arrott-Noakes plots:

$$(H/M)^{1/\gamma} = (T-T_C)/T_1 + (M/M_1)^{1/\beta}, \quad (1)$$

where $T_1$ and $M_1$ are the parameters; $\beta$ and $\gamma$ are the critical indices; $M$ is the magnetization; $H$ is the magnetic field; $T_C$ is the critical temperature.

The MCE was estimated in terms of magnetic entropy change $-\Delta S_M$ from isothermal curves $M(H, T)$ in a wide temperature range of $[T_C - 40$ K; $T_C + 40$ K$]$ with a step of $\Delta T = 2$ K and in field $H$ up to 3 T using the numerical integration method of the Maxwell relation [3]:

$$\Delta S_M(T,H) = S_M(T,H) - S_M(T,0) = \int_0^H \left(\frac{\partial M}{\partial T}\right)_H dH. \quad (2)$$

## 3. Results and discussion

*3.1. Structural properties*

X-ray diffraction (XRD) patterns of the La$_{0.8-x}$□$_x$Na$_{0.2}$Mn$_{1+x}$O$_{3-\Delta}$ nanopowders refined with the Rietveld method and their structure are shown in Fig. 1a-b. All La$_{0.8-x}$□$_x$Na$_{0.2}$Mn$_{1+x}$O$_{3-\Delta}$ are well described in the framework of rhombohedral space group (SG) $R\overline{3}c$ (No. 167) comparable with ICSD (01-080-3121) [47]. The concentration dependence of lattice parameters for the La$_{0.8-x}$□$_x$Na$_{0.2}$Mn$_{1+x}$O$_{3-\Delta}$ is shown in Supplementary Material (SM) Section S1 (see Table S1). With an increase in both overstoichiometric manganese content and concentration of cation vacancies on *A*-position V$_A^{(c)}$, an additional ferrimagnetic Mn$_3$O$_4$ second-phase with a tetragonal SG I4$_1$/amd (No. 141) comparable with ICSD (01-071-6262) [48] appears and grows. This demonstrates the existence of a solubility limit for overstoichiometric manganese up to $x < 0.10$ in the La$_{0.8-x}$□$_x$Na$_{0.2}$Mn$_{1+x}$O$_{3-\Delta}$ nanopowder. The similar results have been observed earlier [18, 21, 25, 26, 33], where, on the one hand, the creation of only cation vacancies V$_A^{(c)} \geq 0.09$ already leads to the appearance of Mn$_3$O$_4$ second-phase and, on the other hand, at the concentration rate of overstoichiometric manganese $\leq 0.10$, the manganite system is still single-phase.

According to Transmission Electron Microscope (TEM) images for the La$_{0.8-x}$□$_x$Na$_{0.2}$Mn$_{1+x}$O$_{3-\Delta}$ nanopowders (see Figs. 1c-g), an average size of spherical-like particles *d* (see S2) are $d = 73$ nm ($x = 0.00$), 55 nm ($x = 0.05$), 48 nm ($x = 0.10$), 62 nm ($x = 0.15$) and 69 nm ($x = 0.20$) that correlates with the coherent scattering region *D* (at $x \leq 0.10$) based on the XRD data (see S1). Discrepancies between TEM and XRD data for other samples with increased content of overstoichiometric manganese $x > 0.10$ can be associated with Mn$_3$O$_4$ second-phase and with the difficulty of determining the true size due to a magnetic attraction between the nanoparticles. The high-resolution TEM (HRTEM) image clearly shows the lattice interplanar spacing of well crystalline nanoparticles with an average size 0.381 nm determined from the lattice plane intensity profile corresponding to (012) plane (see Fig. 1h).

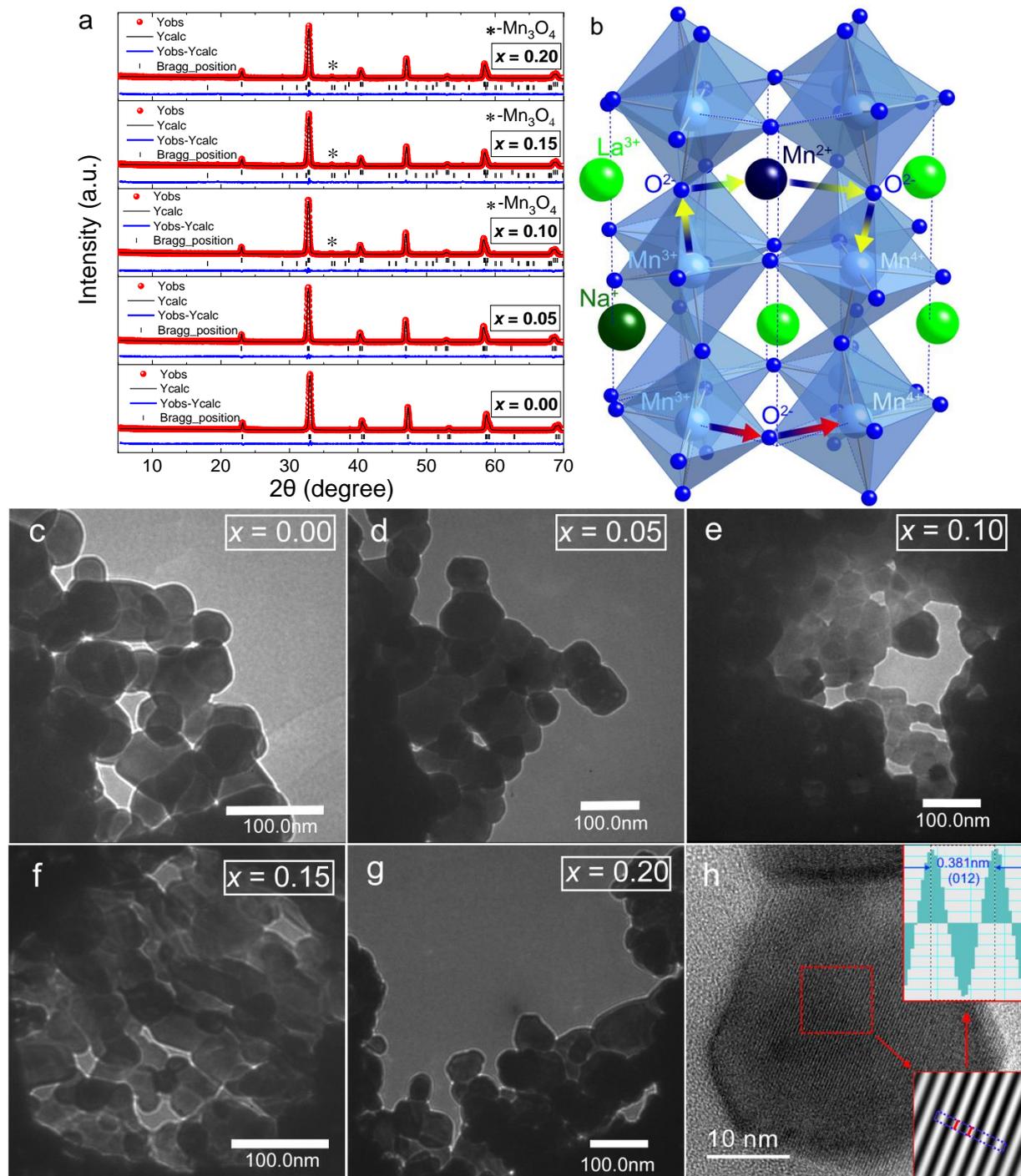

**Fig. 1.** Evolution of XRD patterns and morphology in the $La_{0.8-x}\square_{x}Na_{0.2}Mn_{1+x}O_{3-\Delta}$ nanoparticles with a perovskite structure: (a) XRD patterns measured at room temperature and fitted by Rietveld method show the preserving rhombohedral $R\bar{3}c$ structure (upper bars) and the nucleation of hausmannite $Mn_3O_4$ second-phase (down bars) as both the overstoichiometric manganese content and the concentration of cation vacancies on $A$-position $V_A^{(c)}$ increase; (b) The general scheme of unit cell shows the location of ($La^{3+}$, $Na^+$, $Mn^{2+}$) cations on $A$-, ($Mn^{3+,4+}$) cations on $B$- and anions ($O^{2-}$) on $X$-sites of the perovskite structure with possible mechanisms of DE $Mn^{3+}\leftrightarrow O^{2-}\leftrightarrow Mn^{4+}$ and multiple DE $Mn^{3+}\leftrightarrow O^{2-}\leftrightarrow Mn^{2+}\leftrightarrow O^{2-}\leftrightarrow Mn^{4+}$; (c-g) TEM images show a non-monotonic change in the size of spherical-like nanoparticles with increase in $x$; (h) HRTEM image indicates well-developed crystalline structure of nanoparticles with 0.381 nm corresponding to (012) plane determined from lattice plane intensity profile using the fast Fourier transformation.

According to the established defect formation mechanism for real perovskite structure with point defects of vacancy type (cation $V^{(c)}$ and anion $V^{(a)}$ vacancies) [26, 49, 50], the overstoichiometric manganese can occupy the cation deficient *A*- and *B*-sites in form of $Mn_A^{2+}$ and $Mn_B^{3+}/Mn_B^{4+}$ ions, respectively. The appearance of $Mn^{2+}$ ions was observed earlier for similar *A*-cation deficient manganites [17, 18, 20, 21] and will be proved below using (X-ray photoelectron spectroscopy) XPS method. The $Mn^{2+}$ ions with effective ionic radius 1.13 Å for coordination number (CN) = 12 occupy the *A*-site along with the bigger $La^{3+}$ (1.36 Å) and $Na^+$ (1.39 Å) ions (see Fig. 1b) that leads to distortion of the perovskite structure [19, 20, 51]. The availability of $Mn^{2+}$ ions in real defect manganites can additionally be associated with the charge disproportionation model [52], i.e. the appearance of the more stable $Mn^{2+} + Mn^{4+}$ pairs instead of $Mn^{3+} + Mn^{3+}$. Moreover, it has been reported recently that $Mn^{2+}$ ions can form on the *A*-sites at the reconstructed surface up to ~ 7 nm of the $LaMnO_3$ single crystal particles due to La-deficient particle surface layers [53].

As shown in Table S1, the unit cell volume tends to decreasing with increase in *x*. The similar behavior was reported for *A*-deficient manganites [18, 22, 24]. One of the interesting features obtained from structural data is the change in bond distance and angle depending on the *x* which play crucial role in magneto-transport properties of manganites (see Fig. 1b). To establish regularities between structural and magnetic (see below) properties of the $La_{0.8-x}\square_xNa_{0.2}Mn_{1+x}O_{3-\Delta}$, we defined the bond distances $<d_{Mn-O}>$ for $Mn_B^{3+/4+} - O^{2-}$ / $La_A^{3+}/Na_A^+/Mn_A^{2+} - O^{2-}$ and angles Mn−O−Mn for $Mn_B^{3+/4+} - O^{2-} - Mn_B^{3+/4+}$ / $La_A^{3+}/Na_A^+/Mn_A^{2+} - O^{2-} - Mn_B^{3+/4+}$ between corresponding positions of Mn ions on *A*- and *B*-sites and oxygen taking into account the iodometric titration (IT) data with the certain concentration of $V^{(a)}$ and excluding influence of $Mn_3O_4$ second-phase. As *x* increases, the both lattice parameters and $<d_{Mn-O}>$ ($La_A^{3+}/Na_A^+/Mn_A^{2+} - O^{2-}$) decrease testifying to an appearance of $Mn_A^{2+}$ [19, 20]. A trend in increasing $<d_{Mn-O}>$ ($Mn_B^{3+/4+} - O^{2-}$) and decreasing Mn−O−Mn ($Mn_B^{3+/4+} - O^{2-} - Mn_B^{3+/4+}$) should lead to reducing bandwidth $W \sim \cos\theta \cdot <d_{Mn-O}>^{-3.5}$, weakness of the DE and, subsequently, decreasing the $T_C$ [41]. However, on the other hand, the $<d_{Mn-O}>$ for $Mn_B^{3+/4+} - O^{2-}$ (at small $x < 0.10$) and $La_A^{3+}/Na_A^+/Mn_A^{2+} - O^{2-}$ decreases and the Mn−O−Mn for $La_A^{3+}/Na_A^+/Mn_A^{2+} - O^{2-} - Mn_B^{3+/4+}$ increases which favorable affects the magneto-transport properties of manganites owing to the $Mn_A^{2+}$ ions participating in the multiple DE hopping [18-20]. Thus, the different mechanisms of exchange interaction and their strength will define the magnetic behavior of the $La_{0.8-x}\square_xNa_{0.2}Mn_{1+x}O_{3-\Delta}$ (see below).

For determining the surface chemical composition and valence state of ions, XPS method was employed. Figure 2a shows Mn2p spectra and their decomposition into three components that confirms availability of different valence $Mn^{2+}$, $Mn^{3+}$ and $Mn^{4+}$ ions in the $La_{0.8-x}\square_xNa_{0.2}Mn_{1+x}O_{3-\Delta}$ and even for the pristine composition with $x = 0.00$. The other La3d, Na1s and O1s spectra with their energy positions and valence states are presented in S3. It should be noted that La and Na ions are in the 3+ and 1+ oxidation states, respectively. Additionally, the presented decomposition results (see Table S6) were used to determine the chemical composition of the samples containing Mn ions with different charges taking into account the electroneutrality principle [54]. Figures 2b-d show the procedure for finding element concentrations in the $La_{0.8-x}\square_xNa_{0.2}Mn_{1+x}O_{3-\Delta}$ at different intensities of the analytical O1s line corresponding 1, 2 and 3 points (see details in S3). The requirement for the electroneutrality sample is satisfied upon the special value of oxygen non-stoichiometry Δ when charge balance = 0 (point *N* in Figs. 2b-d). The vertical dashed line intersects La, Na, Mn and O plots through the point *N* and gives the element concentrations for the electrically neutral $La_{0.8-x}\square_xNa_{0.2}Mn_{1+x}O_{3-\Delta}$. The obtained XPS results for the concentration of ions and oxygen content are consistent with IT (see Table S1) and (Energy dispersive X-ray spectroscopy) EDS data (see Table S7), which were additionally employed for finding chemical composition taking into account electroneutrality principle and based on the non-defect perovskite structure (see Table S8).

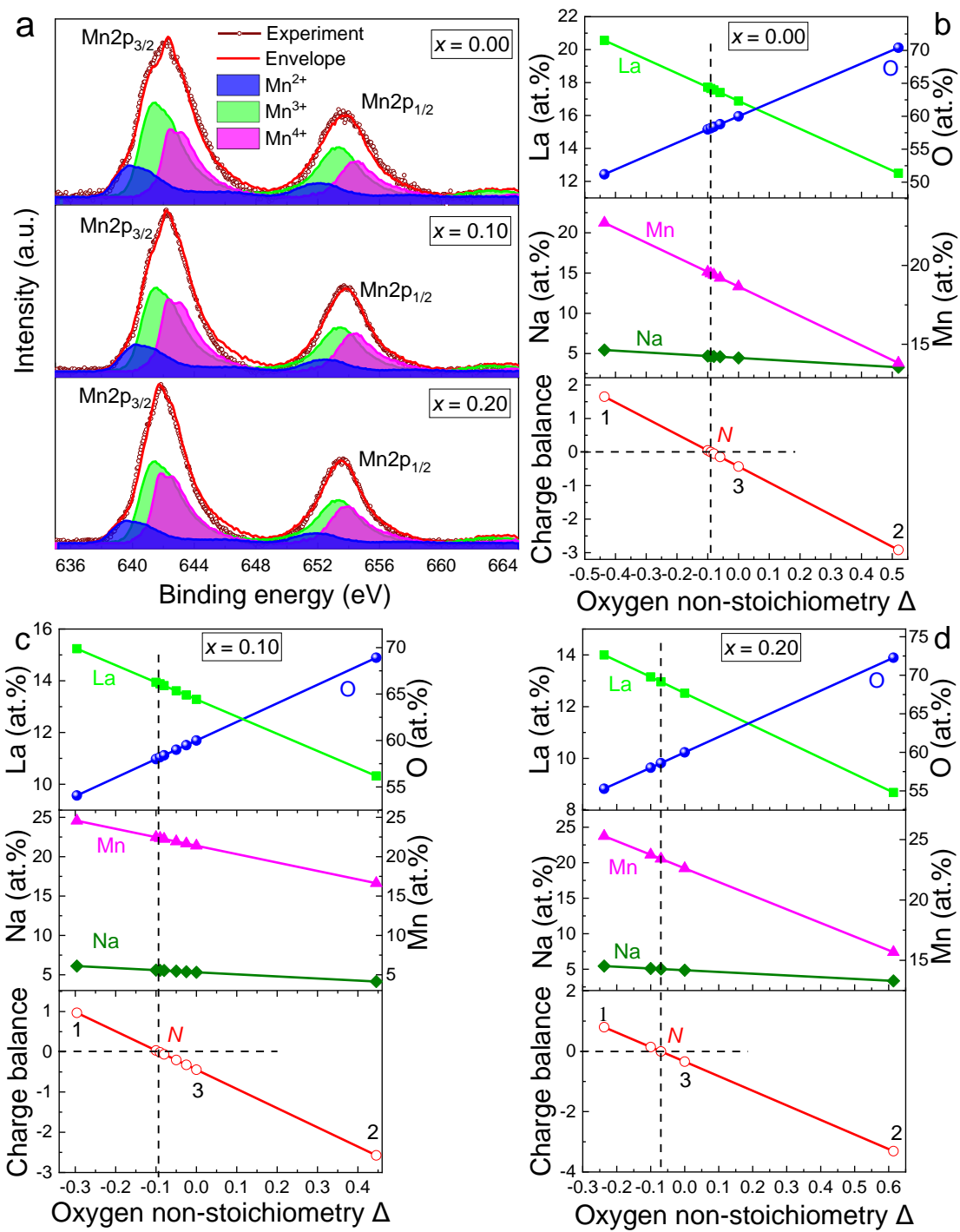

**Fig. 2.** Mn2p spectra and refinement procedure of chemical composition for the $La_{0.8-x}\square_x Na_{0.2}Mn_{1+x}O_{3-\Delta}$: (a) The Mn2p spectra approximated by red envelope line show an availability of different valence state of $Mn^{2+}$, $Mn^{3+}$ and $Mn^{4+}$ ions in all samples and even in the pristine $La_{0.8}Na_{0.2}MnO_3$; (b-d) Elemental concentrations determined at various intensities of the analytical O1s line (points 1, 2, 3) and according to electroneutrality principle (charge balance = 0 in point $N$) indicate their content in the $La_{0.8-x}\square_x Na_{0.2}Mn_{1+x}O_{3-\Delta}$, which correlate with IT and EDS data (see Table S8).

Thus, based on the analysis of XRD, XPS, IT, EDS, TEM and HRTEM data, there are $La^{3+}$, $Na^+$ cations and oxygen deviation $O_{3-\Delta}$, as well as different valence $Mn^{2+}$, $Mn^{3+}$ and $Mn^{4+}$ ions which play an important role in the formation of magnetic and magnetocaloric properties of the spherical-like nanoparticles of non-stoichiometric $La_{0.8-x}\square_xNa_{0.2}Mn_{1+x}O_{3-\Delta}$ manganite. It should be noted that in the next sections, we will consider only compositions with low concentration level of overstoichiometric manganese ($x \leq 0.10$) and the most structural and magnetic inhomogeneity ($x = 0.20$).

*3.2. Spin-dependent magnetism of manganese ions*

During transition from FM state to PM state, we have found an interesting result: the magnetic moment of Mn ions changes significantly towards its increasing (see Fig. 3). In the FM phase, the electronic configuration of $Mn^{2+}$, $Mn^{3+}$, $Mn^{4+}$ ions being in the high-spin state is $3d^5$ ($e_g^{2\uparrow}t_{2g}^{3\uparrow}$) with spin $S = 5/2$ and magnetic moment $\mu = 5\mu_B$, $3d^4$ ($e_g^{1\uparrow}t_{2g}^{3\uparrow}$) with $S = 2$ and $\mu = 4\mu_B$, $3d^3$ ($t_{2g}^{3\uparrow}$) with $S = 3/2$ and $\mu = 3\mu_B$, respectively. Using the saturation magnetization $M_S$ values at 2 K (see Fig. S3 and Table S9 in S4), the experimental magnetic moment $\mu_{FM}^{exp}$ can be obtained for the $La_{0.8-x}\square_xNa_{0.2}Mn_{1+x}O_{3-\Delta}$ (see Fig. 3 and Table S10). At $x \leq 0.10$, the $\mu_{FM}^{exp}$ is in good agreement with the theoretical values based on the $\mu_{FM}^{theor} = g\mu_B S$ with Landé factor $g = 2$ and IT data (see Table S8). However, as $x$ increases the magnetic moments have the different tendencies, i.e. the $\mu_{FM}^{exp}$ and $\mu_{FM}^{theor}$ decreases and increases, respectively.

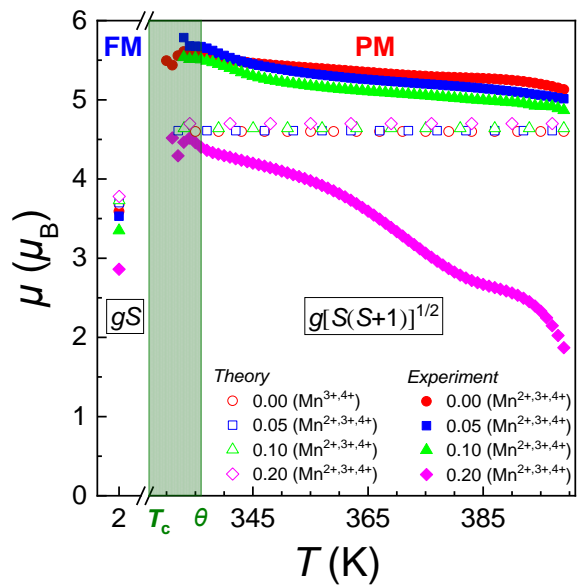

**Fig. 3.** Temperature dependence of the magnetic moment $\mu(T)$ for the $La_{0.8-x}\square_xNa_{0.2}Mn_{1+x}O_{3-\Delta}$ being in FM and PM states. With increase in $T$ during transition from FM to PM state through the Curie $T_C$ and paramagnetic Curie $\theta$ temperature region, there are: (i) a significant leap in the experimental magnetic

moments ($\mu_{FM}^{exp} \to \mu_{PM}^{exp} \uparrow$); (ii) a large deviation of experimental $\mu_{PM}^{exp}$ from theoretical $\mu_{PM}^{theor}$; (iii) an appearance of overstated content of PM $Mn^{2+}$ ions.

In PM state at $T > T_C$ (see Fig. S4) using the Curie-Weiss law $\chi = C / (T - \theta)$ with the PM Curie temperature $\theta$, the Curie constant $C = N(\mu_{PM})^2/(3k_B)$, the number of PM manganese ions $N$ and the Boltzmann constant $k_B$, we can obtain the effective magnetic moment of manganese $\mu_{PM}$ in Bohr magnetons $\mu_B$ (see Fig. S5 and Table S10) that has an opposite situation compared with FM state. The experimental $\mu_{PM}^{exp}$ significantly exceed the values of theoretical effective magnetic moment $\mu_{PM}^{theor} = g\mu_B[S(S+1)]^{1/2}$ taking into account the quenched orbital angular momentum by the crystalline field for $Mn^{3+}$ and $Mn^{4+}$ ions being in the high-spin state with $S = 2$ and $S = 3/2$ within a model of localized electrons, i.e. $\mu_{Mn}^{3+} = 4.90\mu_B$ and $\mu_{Mn}^{4+} = 3.87\mu_B$ [55]. This indicates the presence of $Mn^{2+}$ ions with $\mu_{Mn}^{2+} = 5.748\mu_B$ [55] in all compositions and even in the pristine $La_{0.8}Na_{0.2}MnO_{3-\Delta}$ (see Fig. 3 and Table S10). The availability of $Mn^{2+}$ ions can be associated with their appearance on the surface of nanoparticles for all compositions on the $A$-deficient sites instead of La ions with reduced CN [53] that is in good agreement with our XPS data (see Table S8). Moreover, we defined the contribution from the shell containing $Mn^{2+}$ ions in the $\mu_{PM}^{exp}$ (see S4) based on the XPS and IT data (see Table S8). It turned out that for the spherical-like nanoparticle with a size of 61 nm (see Table S1 for $x = 0.00$) the contribution from the surface layer in the $\mu_{PM}^{exp}$ is about of 9% that gives $0.537\mu_B$ and consequently $4.693\mu_B$ for the core containing only $Mn^{3+/4+}$ ions. For all other concentrations $x$ besides the surface layer containing $Mn^{2+}$ ions (see Table S8), the contribution will be also defined from $Mn^{2+}$ ions being inside the perovskite structure which is confirmed by decreasing lattice parameters (see Table S1).

Thus, during transition from FM state to PM state, the significant changes in the experimental magnetic moments ($\mu_{FM}^{exp}$ and $\mu_{PM}^{exp}$) as well as their deviation from theoretical ones ($\mu_{FM}^{theor}$ and $\mu_{PM}^{theor}$) are observed depending on $x$. This situation can be caused by different mechanisms, such as: (i) an existence of short-range FM clusters in PM phase [56]; (ii) an appearance of the low spin-state of $Mn^{2+}$ ions in the tetrahedral $B$-sites [57]; (iii) a change spin value of Mn ions depending on the temperature and, consequently, the exchange interaction energy [14, 40, 58, 59]. The (i) mechanism can exist in the range from $T_C$ to $\theta$ [56] that for our compositions achieves maximum value 5 K difference (see Table S10) and cannot significantly influence the experimental magnetic moments $\mu_{PM}^{exp}$ above this range. The (ii) mechanism may occur under extreme conditions, for example, because of reduced oxygen content to "$O_{2.5}$" in manganites [57] that does not consistent with our XPS, EDS and IT data (see Table S8). In our opinion, the most probable mechanism is (iii). As a rule for manganites, near FM-PM transition, there is a change in the nature of conductivity from metallic to semiconductor with a localization of $e_g$-electrons [40]. At the same time, the distribution of the charge density occurs according to the rule of local and overall charge preservation. As a result, a variation in the valence and the spin state of the manganese ions is observed which leads to a change in the overlap of the wave functions of manganese and oxygen ions and, consequently, to alteration of exchange interactions (see S5). With increase in temperature and at the transition *via* $T_C$, the magnetic moment of Mn increases due to changing its spin value with a localization of $e_g$-electrons on the PM manganese ions. With a reduce in temperature and upon transition from PM to FM state, the magnetic moment of Mn ions decreases dramatically because of possible changing exchange interaction energy between manganese spins. As *x* increases the experimental magnetic moment for the both FM and PM states is decreased that can be associated with the charge disproportionation model [52], i.e. an appearance of $Mn^{4+}$ ions and increase in their concentration (see Table S8) with lower spin $S = 3/2$.

Therefore, based on the XPS spectroscopy, IT and magnetic measurements data, there are differences in the valence, charge and spin state of Mn ions in FM and PM regions. These differences lead to the change of the spin value (not only of its projection) in the FM and PM states and, consequently, the magnetic moments of Mn ions. It can be concluded that the spin-dependent effect of changing magnetic moment of Mn ions in PM and FM states are observed. In this case, the magnetic behavior will be determined by the spin-dependent exchange. Such type of nonlinearity of magnetism in the manganites should strongly affect their behavior in the critical region.

*3.3. Critical magnetization*

Within temperature range $T$ = 280–371 K, the $M(H)$ curves were measured for all $La_{0.8-x}\square_xNa_{0.2}Mn_{1+x}O_{3-\Delta}$ (see S6). At $T > T_C$, the magnetization is proportionally to the field $M \sim H$ and the particles are in the PM state. At $T < T_C$, the magnetization reaches saturation nonlinearly with Langevin behavior [60] that is typical for the SPM state [32].

To analyze the nature of the magnetic phase transition for the $La_{0.8-x}\square_xNa_{0.2}Mn_{1+x}O_{3-\Delta}$, the critical indices near the critical temperature should be found. For the second-order phase transition and according to the scale hypothesis, the critical behavior near $T_C$ is characterized by a set of critical exponents $\delta$, $\beta$, $\gamma$ associated with the critical isotherm magnetization, the spontaneous magnetization and the isothermal magnetic susceptibility, respectively [39].

The dependences of the logarithm of the magnetization, $\ln(M)$, *versus* the logarithm of the magnetic field, $\ln(H)$, for the $La_{0.8}\square_xNa_{0.2}Mn_{1+x}O_{3-\Delta}$ are shown in Fig. S6, which allow us to find the critical index $\delta$ from the power law dependence $M \approx aH^{1/\delta}$ (see Table S11), where $a$ is the constant [61].

In the strong field, the asymptote of magnetization $M$ for the SPM particles is $(M_S - M) \sim H^1$ [37]. It allows us to find the temperature dependence of the spontaneous magnetization $M_S(T)$ for all $x$ (see Fig. S7), and then to determine critical index $\beta$ from the power law dependence $M_S(T) \sim (T-T_0)^\beta$ [37]. According to the $M_S^{1/\beta}(T)$ dependence (see the insets in Fig. S7), the critical indices $\beta$ and critical temperatures $T_0$ were determined (see Table S11).

To better understand the internal magnetic interactions between spins, the critical exponents should be analyzed by Arrott-Noakes plots $(M)^{1/\beta} = f(H/M)^{1/\gamma}$ [1, 62, 63] using Eq. (1). The Arrott plots with using different models, such as: (i) Mean field model (MFM) with $\beta$ = 0.5, $\gamma$ = 1.0; (ii) Tricritical mean field model (TMFM) with $\beta$ = 0.25, $\gamma$ = 1.0; (iii) 3D-Ising model with $\beta$ = 0.325, $\gamma$ = 1.24; (iv) 3D-Heisenberg model with $\beta$ = 0.365, $\gamma$ = 1.336; (v) Modified Arrott plot (MAP) model with $\delta$ = 3.20, $\beta$ = 0.37 and $\gamma$ = 0.814 determined from the Windom scaling law $\delta$ = 1+ $\gamma/\beta$ [64] are shown in Fig. 4 for only $x$ = 0.00. The values of all critical $\delta$, $\beta$, $\gamma$ indices with the defined critical $T_{MFM}$, $T_{TMFM}$, $T_{Ising}$, $T_{\text{Heisenberg}}$ and $T_{\text{MAP}}$ temperatures depending the model and their corresponding Arrott plots for all *x* are presented in Table S11 and Figs. S8a-x, respectively. In the high field region, all five models show good parallel behavior. To compare these results and determine the best model for describing magnetic phase transition, their relative slopes (RS) defined as RS = $S(T)/S(T_C)$ are used, where *S* is the slope of model curve in the high magnetic region. As turned out for all *x*, the most suitable model is MAP model with δ, β and γ determined from the experimental procedure testifying to the strong nonlinear spin exchange in the $La_{0.8-x}\square_x Na_{0.2}Mn_{1+x}O_{3-\Delta}$ nanoparticles due to coexistence of traditional and multiple DE (see Fig. 1b) near $T_C$ in compounds with overstoichiometric manganese [19, 20]. The MAP model is very close to "1" near critical temperature $T_{\text{MAP}}$ (see Figs. S8a-x) and perfectly coincides with the $T_C$ (see Table S11). Moreover, according to Banerjee criterion [65], the $La_{0.8}\square_x Na_{0.2}Mn_{1+x}O_{3-\Delta}$ demonstrate the second-order phase transition with the positive slope of isotherm $M^2(H/M)$ (see Figs. S8a-x) indicating validity of all used models.

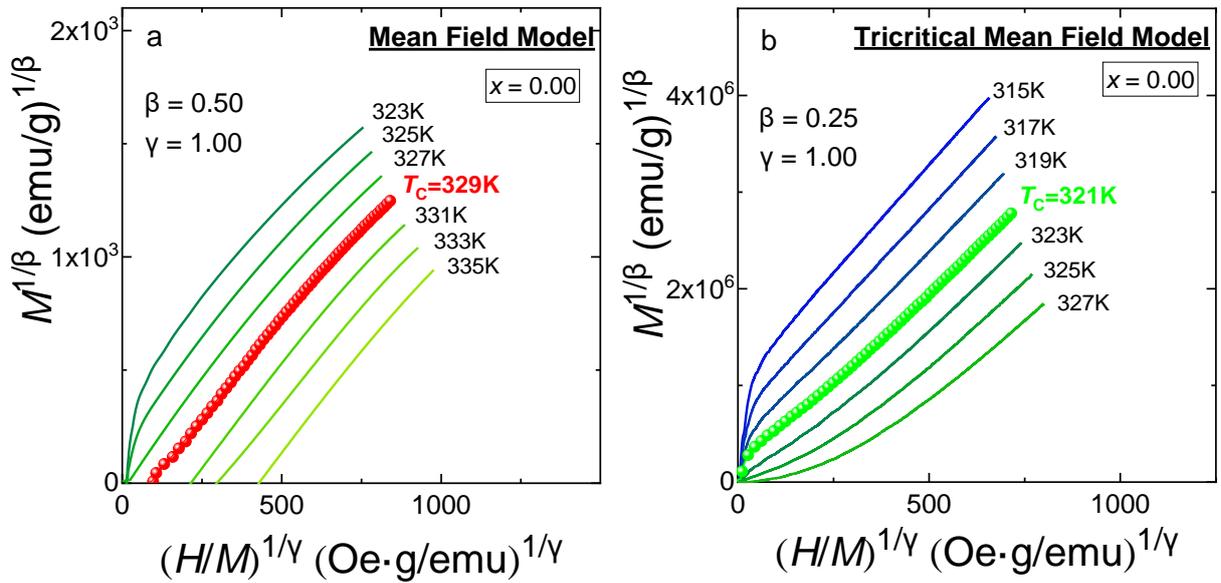

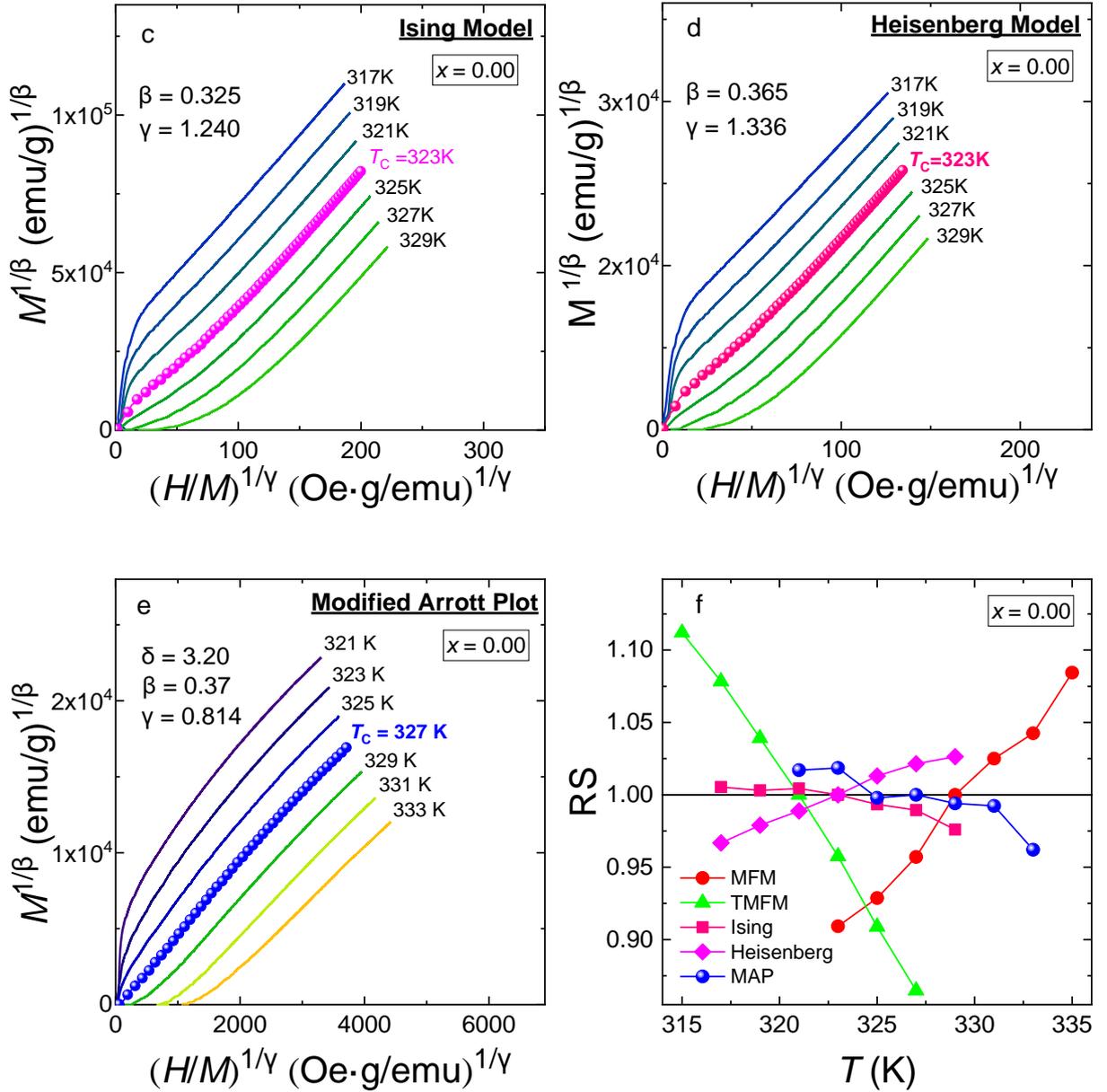

**Fig. 4.** Arrott plots for the $La_{0.8-x}\square_x Na_{0.2}Mn_{1+x}O_{3-\Delta}$ with $x = 0.00$: (a-f) Among different approaches, the modified Arrott plot model demonstrates the best describing magnetic phase transition near critical temperature based on the relative slope $RS = S(T)/S(T_C)$.

Thus, the critical temperature obtained from the MAP model can be additionally found from the magnetic entropy change $-\Delta S_M$, which has a feature in the critical point $T_{MCE}$. In such non-Heisenberg $La_{0.8-x}\square_x Na_{0.2}Mn_{1+x}O_{3-\Delta}$ system, the magnetic moments of Mn ions are ordered by the both exchange Heisenberg and non-Heisenberg fields which should lead to enhance of the MCE near critical point.

*3.4. Magnetocaloric effect and its features near critical point*

The magnetic entropy change $-\Delta S_M(T, H)$ for the La$_{0.8-x}\square_x$Na$_{0.2}$Mn$_{1+x}$O$_{3-\Delta}$ (see Fig. 5 and S7) were plotted using Eq. (2). The maximum of $|-\Delta S_M^{max}|$ and its critical temperature $T_{MCE}$ increase for $x \leq 0.10$, and then dramatically reduce for $x = 0.20$. At the same time, a big difference in the values of $T_{MCE}$ under low and high magnetic field is observed that is associated with SPM of nanoparticles (see Fig. 5 and Fig. S10). Indeed, at small $H \leq 2$ kOe, SPM contribution in the largest change of $|-\Delta S_M(H)|$ occurs at lower temperatures $T_{MCE}$ ($H = 2$ kOe) $< T_{MCE}$ ($H = 30$ kOe) (see Fig. S11).

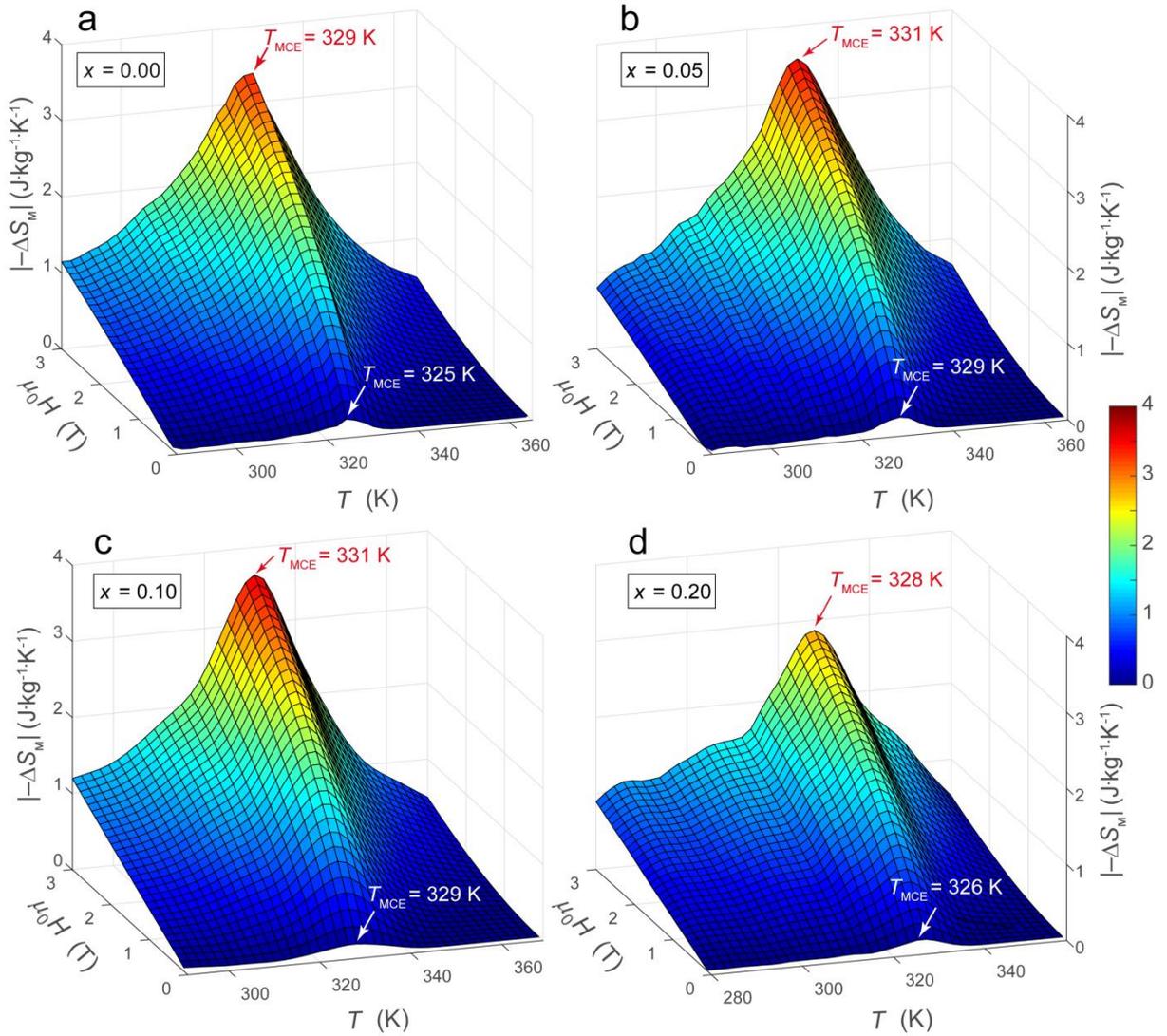

**Fig. 5.** Evolution of magnetic entropy change for the La$_{0.8-x}\square_x$Na$_{0.2}$Mn$_{1+x}$O$_{3-\Delta}$ nanopowders: (a-d) With increase in magnetic field, the MCE temperature $T_{MCE}$ is shifted toward a high temperature region for all compounds and can achieve $\Delta T_{MCE} = 4$ K. Among others, the La$_{0.7}\square_{0.1}$Na$_{0.2}$Mn$_{1.1}$O$_{3-\Delta}$ shows the highest $|-\Delta S_M|$.

Additionally, Fig. 6 shows the field dependences of derivative of $-d\Delta S_M(T = \text{const}, H)/dH$, which achieves maximum values under 0.30 kOe at $T_{MCE} = 325$ K ($x = 0.00$), 0.66 kOe at $T_{MCE} = 329$ K ($x = 0.05$), 0.46 kOe at $T_{MCE} = 329$ K ($x = 0.10$) and 0.16 kOe at $T_{MCE} = 326$ K ($x = 0.20$) for particles being in the SPM state that is more than 2 times higher $|-d\Delta S_M(T = \text{const}, H)/dH|$ values for particles being in PM state ($T > T_{MCE}$). At lower $T_{MCE}$, the rate of entropy change increases linearly as $|-d\Delta S_M(T = \text{const}, H)/dH| \sim H$ up to the maximum point and then slowly decreases with an increase in $H$. In a stronger magnetic field $H > 2$ kOe, the difference between SPM and PM contributions in the magnetic entropy becomes insignificant. With further increase in $H$, the SPM contribution caused by maximally disoriented magnetic moments along the field disappears and all particles go to FM state, but at higher temperatures of $T_{MCE}$.

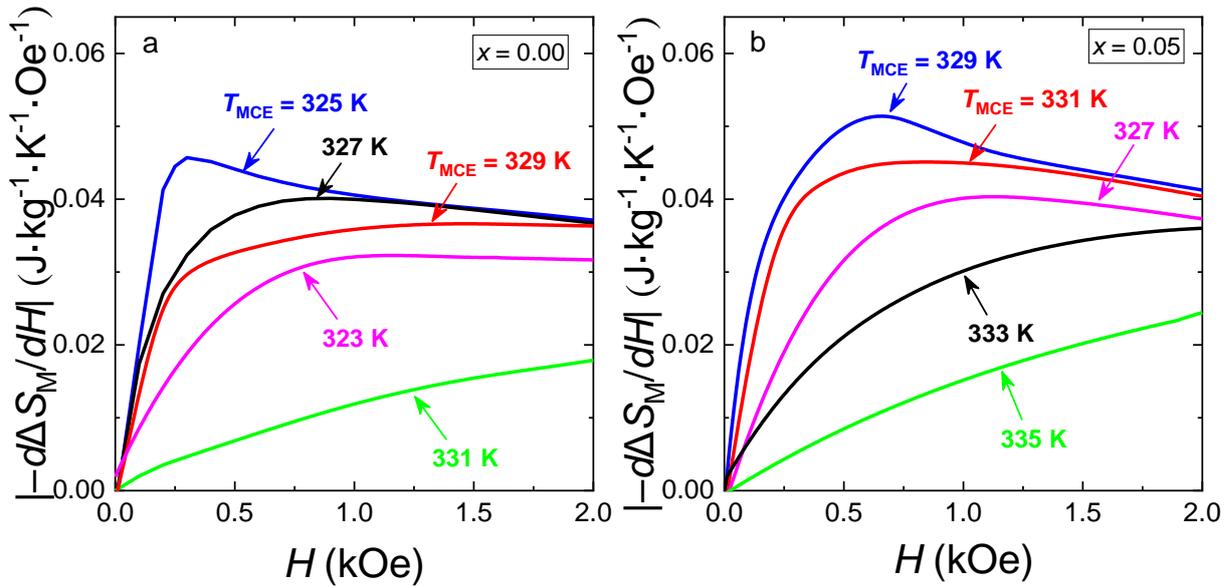

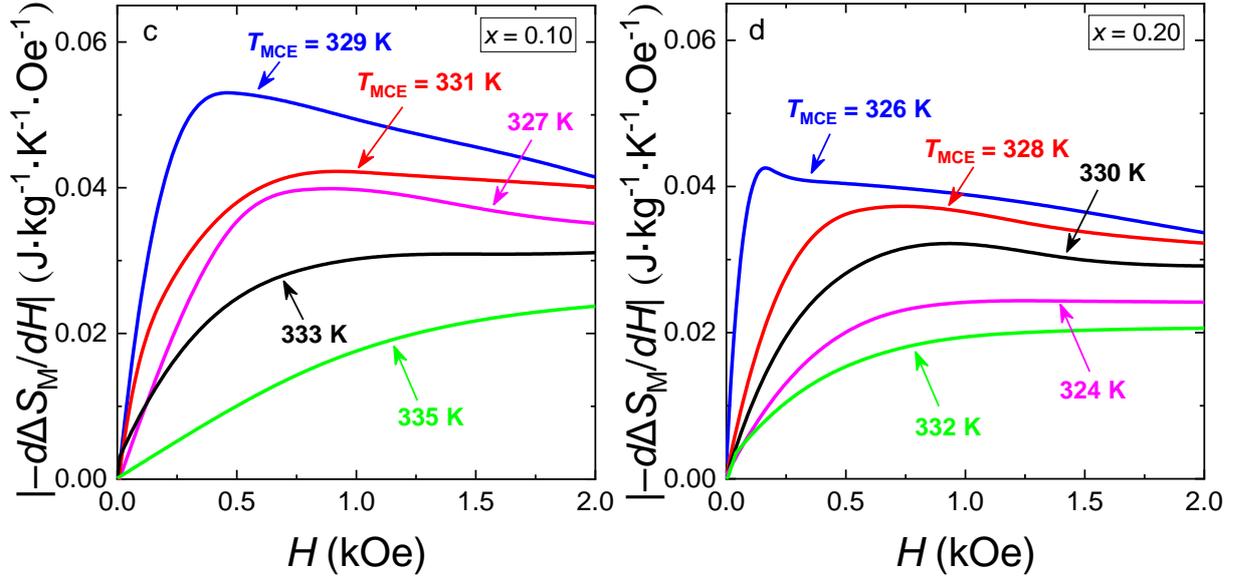

**Fig. 6.** Field dependences of derivative of the magnetic entropy change for the $La_{0.8-x}\square_xNa_{0.2}Mn_{1+x}O_{3-\Delta}$: (a-d) In the small magnetic field, the peak of $|-d\Delta S_M(T=const, H)/dH|$ is observed at lower $T_{MCE}$ due to SPM of nanoparticles.

## 4. Conclusions

In this paper, we have designed the nonlinear $La_{0.8-x}\square_xNa_{0.2}Mn_{1+x}O_{3-\Delta}$ nanoparticles and studied the influence of chemistry defects and overstoichiometric manganese on their structure, morphology, valence, charge and spin states of ions, as well as magnetic and magnetocaloric properties. It has been found that all compounds consist of $Mn^{2+}$, $Mn^{3+}$, $Mn^{4+}$ ions, the magnetic moment of which and their exchange are changed depending on the concentration, temperature and magnetic state. During transition from the FM state to the PM state and *vice versa*, the magnetic moment of Mn ions and, as a consequence, their spin changes significantly because of alteration in (i) movability of $e_g$-electrons (delocalization/localization) and (ii) charge density distribution pointing to the existence of spin-dependent magnetism. It has been shown that magnetic behavior of non-Heisenberg $La_{0.8-x}\square_xNa_{0.2}Mn_{1+x}O_{3-\Delta}$ samples is perfectly described within the MAP model with the second order magnetic phase transition near $T_C$. Among all compositions, the $La_{0.7}Na_{0.2}Mn_{1.1}O_{3-\Delta}$ nanopowder exhibits the highest $|-\Delta S_M^{max}|$ near room temperature. It has been discovered that the temperature of magnetic entropy change peak is shifted towards higher temperatures with increase in magnetic field. Such anomaly is caused by two factors: (i) at small field, the main contribution in MCE originates from the SPM of nanoparticles with maximally disoriented magnetic moments; (ii) at high field, the main contribution in MCE is already determined by the transition of nanoparticles to FM state with the field aligned magnetic moments. The presented studies and obtained results show the perspectives of the creation of chemistry defects and implementation of overstoichiometric manganese as $Mn_A^{2+}$ ions in the perovskite structure of manganites in order to tune their magnetic phase transition temperature and magnetocaloric parameters.

**Authors' contribution**

All authors were involved in the conception and design of the work. N.A.L., V.M.K., A.V.P., A.T.K., and G.G.L. extensively discussed the results and commented on the manuscript. N.A.L., Z.W., V.M.K., Q.L., I.V.F., V.A.T., C.H., and A.T.K. participated in performing experiments, performed the calculations and analysis. N.A.L. wrote the manuscript. N.A.L. and Z.W. contributed equally to the manuscript. N.A.L., A.T.K., and G.G.L. led the project.

**Declaration of competing interest**

The authors declare that they have no known competing financial interests or personal relationships that could have appeared to influence the work reported in this paper.

**Acknowledgements**

This work was supported by The Thousand Talents Program for Foreign Experts program of China (project WQ20162200339), Grant of NAS of Ukraine for research laboratories / groups of young scientists of NAS of Ukraine 2020–2021 (№ 08/01-2021) and Internal grant of Southern Federal University for the implementation of scientific research (project No. VnGr -07/ 2020-01-IF).

**Data Statement**

All data that support the plots and other findings within this paper are available from the corresponding authors on reasonable request. Source data are provided with this paper.

# Supplementary Material

# Spin-dependent magnetism and superparamagnetic contribution to the magnetocaloric effect of non-stoichiometric manganite nanoparticles

Nikita A. Liedienov [a, b], Ziyu Wei [a], Viktor M. Kalita [c, d, e], Aleksey V. Pashchenko [a, b, d, *], Quanjun Li [a], Igor V. Fesych [f], Vitaliy A. Turchenko [g, h], Changmin Hou [i], Aleksey T. Kozakov [j], Georgiy G. Levchenko [a, b, *]

[a] *State Key Laboratory of Superhard Materials, International Center of Future Science, Jilin University, 130012 Changchun, P. R. China*
[b] *Donetsk Institute for Physics and Engineering named after O.O. Galkin, NAS of Ukraine, 03028 Kyiv, Ukraine*
[c] *National Technical University of Ukraine "Igor Sikorsky Kyiv Polytechnic Institute", 03056 Kyiv, Ukraine*
[d] *Institute of Magnetism, NAS of Ukraine and MES of Ukraine, 03142 Kyiv, Ukraine*
[e] *Institute of Physics, NAS of Ukraine, 03028 Kyiv, Ukraine*
[f] *Taras Shevchenko National University of Kyiv, 01030 Kyiv, Ukraine*
[g] *Frank Laboratory of Neutron Physics, Joint Institute for Nuclear Research, 141980 Dubna, Russia*
[h] *South Ural State University, 454080 Chelyabinsk, Russia*
[i] *State Key Laboratory of Inorganic Synthesis and Preparative Chemistry, College of Chemistry, Jilin University, 130012 Changchun, P. R. China*
[j] *Scientific-Research Institute of Physics at Southern Federal University, 344194 Rostov-na-Donu, Russia*

These authors contributed equally: N.A. Liedienov, Ziyu Wei
* Corresponding author
*E-mail addresses*:   alpash@ukr.net (Aleksey V. Pashchenko)
                     g-levch@ukr.net (Georgiy G. Levchenko)

# S1. Structural parameters of the La$_{0.8-x}\square_x$Na$_{0.2}$Mn$_{1+x}$O$_{3-\Delta}$ nanopowders

**Table S1**

Structural parameters of the La$_{0.8-x}\square_x$Na$_{0.2}$Mn$_{1+x}$O$_{3-\Delta}$ compounds refined with the Rietveld method at room temperature: coherent scattering region ($D$); iodometric titration data (O$_{3-\Delta}$); unit cell parameters ($a$, $c$, $V$); atomic positions La/Na(6a) (0, 0, 25), Mn(6b) (0, 0, 0), O(18e) (x, 0, 0); bond distances ($<d_{Mn-O}>$) for $Mn_B^{3+/4+} - O^{2-}$ / $La_A^{3+}/Na_A^+/Mn_A^{2+} - O^{2-}$ and angles (Mn−O−Mn) for $Mn_B^{3+/4+} - O^{2-} - Mn_B^{3+/4+}$ / $La_A^{3+}/Na_A^+/Mn_A^{2+} - O^{2-} - Mn_B^{3+/4+}$ with $R\bar{3}c$ crystal structure and figures of merit ($R_{wp}$, $R_p$, $R_{exp}$, $\chi^2$)

| $x$ | 0.00 | 0.05 | 0.10 | 0.15 | 0.20 |
|---|---|---|---|---|---|
| **Iodometric titration** | | | | | |
| O$_{3-\Delta}$ | 2.96 | 2.96 | 2.95 | 2.95 | 2.93 |
| **Coherent scattering region** | | | | | |
| $D$ (nm) | 61 | 60 | 56 | 55 | 53 |
| **Phase composition** | single-phase | single-phase | ~2 % Mn$_3$O$_4$ | ~5 % Mn$_3$O$_4$ | ~6 % Mn$_3$O$_4$ |
| **Cell parameters** | | | | | |
| $a$ (Å) | 5.4784(1) | 5.4772(1) | 5.4775(1) | 5.4789(1) | 5.4775(1) |
| $c$ (Å) | 13.3166(2) | 13.3183(2) | 13.3179(3) | 13.3214(3) | 13.3178(3) |
| $V$ (Å$^3$) | 346.125(8) | 346.020(8) | 346.046(9) | 346.316(11) | 346.043(11) |
| **Positional parameters** | | | | | |
| O(18e) x | 0.4574(12) | 0.4571(12) | 0.4547(11) | 0.4550(13) | 0.4544(14) |
| **Bond distance $<d_{Mn-O}>$ (Å)** | | | | | |
| $Mn_B^{3+/4+} - O^{2-}$ | 1.9458(50) | 1.9454(50) | 1.9474(46) | 1.9478(55) | 1.9475(59) |
| $La_A^{3+}/Na_A^+/Mn_A^{2+} - O^{2-}$ | 2.5057(66) | 2.5038(66) | 2.4905(60) | 2.4928(71) | 2.4889(77) |
| **Bond angle Mn-O-Mn (°)** | | | | | |
| $Mn_B^{3+/4+} - O^{2-} - Mn_B^{3+/4+}$ | 166.22(22) | 166.13(22) | 165.36(20) | 165.46(24) | 165.27(26) |
| $La_A^{3+}/Na_A^+/Mn_A^{2+} - O^{2-} - Mn_B^{3+/4+}$ | 96.89(15) | 96.94(15) | 97.32(14) | 97.27(16) | 97.37(18) |
| **Agreement factors** | | | | | |
| $R_{wp}$ (%) | 11.8 | 13.2 | 13.8 | 15.8 | 16.2 |
| $R_p$ (%) | 7.27 | 9.49 | 10.00 | 11.3 | 11.2 |
| $R_{exp}$ (%) | 6.67 | 9.39 | 9.59 | 9.52 | 9.51 |
| $\chi^2 = (R_{wp}/R_{exp})^2$ | 2.42 | 1.97 | 2.08 | 2.76 | 2.90 |

The calculation of the X-ray density depending on the oxygen content (O$_{3-\Delta}$) for ideal non-deficient $\rho_{ideal}$ ($\Delta = 0$) and real deficient $\rho_{real}$ ($\Delta \neq 0$) perovskite structure was performed using $\rho = (Z \cdot M_0)/(V \cdot N_A)$, where $Z = 6$ is the number of formula units; $M_0$ is the molar mass of the La$_{0.8-x}\square_x$Na$_{0.2}$Mn$_{1+x}$O$_{3-\Delta}$ composition with different $\Delta$ (see Table S1); $V$ is the unit cell volume (see Table S2); and $N_A = 6.022 \cdot 10^{23}$ mol$^{-1}$ is Avogadro number.

**Table S2**
Phase composition, type of symmetry, lattice parameters ($a$, $c$, $V$), and density ($\rho_{ideal}$, $\rho_{real}$) of the La$_{0.8-x}$□$_x$Na$_{0.2}$Mn$_{1+x}$O$_{3-\Delta}$ ($0 \leq x \leq 0.2$) nanopowder

| $x$ | Space group | Latice parameters | | | Second phase (vol. %) | $\rho_{ideal}$ (g/cm$^3$) | $\rho_{real}$ (g/cm$^3$) |
| --- | --- | --- | --- | --- | --- | --- | --- |
| | | $a$ (Å) | $c$ (Å) | $V$ (Å$^3$) | | | |
| 0.00 | $R\bar{3}c$ | 5.4784(1) | 13.3166(2) | 346.125(8) | - | 6.294 | 6.276 |
| 0.05 | $R\bar{3}c$ | 5.4772(1) | 13.3183(2) | 346.020(8) | - | 6.175 | 6.157 |
| 0.10 | $R\bar{3}c$ | 5.4775(1) | 13.3179(3) | 346.046(9) | 2 | 6.177 | 6.154 |
| 0.15 | $R\bar{3}c$ | 5.4789(1) | 13.3214(3) | 346.316(11) | 5 | 6.240 | 6.216 |
| 0.20 | $R\bar{3}c$ | 5.4775(1) | 13.3178(3) | 346.043(11) | 6 | 6.183 | 6.149 |

The size of the coherent scattering region $D$ was determined using the Debye-Scherrer method. An average particle size $D$ in the La$_{0.8-x}$□$_x$Na$_{0.2}$Mn$_{1+x}$O$_{3-\Delta}$ nanopowders with $x = 0.00$–$0.20$ is related to the dimensional broadening of $\beta$ for diffraction reflection (012) according to the Debye-Scherrer equation [1]:

$$D = K\lambda / \beta \cos\theta, \tag{S1}$$

where $D$ is the size of scattering crystallites in nm; $\lambda = 0.15418$ nm is the wavelength of X-ray radiation; $K \approx 1$ is a constant which depends on the method for determination of the line broadening and crystal shape; $\beta$ is the width of the intensity distribution curve at half maximum reflex in radians; $\theta$ is the diffraction angle in degrees. The true integral width of peak was calculated using the Warren formula [2]: $\beta = \sqrt{\beta_{exp}^2 - \beta_0^2}$, where $\beta_{exp}$ is the total peak width of the measured sample at half maximum intensity in radians; $\beta_0$ is the instrumental broadening of the diffraction line, which depends on the specifications of the diffractometer. The resolution function of the diffractometer $\beta_0$ was determined in other experiment from the diffraction pattern for the cubic hexaboride lanthanum LaB$_6$ standard (NIST Standard Reference Powder 660a).

The average particle size $D$ was obtained using the Lorentz function during approximation of the experimental values of the intensity of the diffraction maximum with Bragg angle of $2\theta \approx 22.9°$ and taking into account all the experimental parameters in Eq. (S1) (see Table S3).

**Table S3**

The experimental parameters in Eq. (S1) and an average size of the coherent scattering region $D$ for the $La_{0.8-x}\square_x Na_{0.2}Mn_{1+x}O_{3-\Delta}$ nanopowders with $x$ = 0.00–0.20

| $x$ | $2\theta$ (degree) | $\beta$ (radian) | $\cos\theta$ | $\lambda$ (nm) | $D$ (nm) |
|---|---|---|---|---|---|
| 0.00 | 22.9080 | 0.0027 | 0.921 | 0.15418 | 61 ±1 |
| 0.05 | 22.9538 | 0.0028 | 0.921 | 0.15418 | 60 ±1 |
| 0.10 | 22.9069 | 0.0030 | 0.921 | 0.15418 | 56 ±1 |
| 0.15 | 22.9986 | 0.0030 | 0.921 | 0.15418 | 55 ±1 |
| 0.20 | 22.9360 | 0.0032 | 0.921 | 0.15418 | 53 ±1 |

## S2. Particle size distribution functions in the La$_{0.8-x}$□$_x$Na$_{0.2}$Mn$_{1+x}$O$_{3-\Delta}$ nanopowders

The type of the distribution function $f = f(D)$ was selected among the three most frequently used functions: the Gauss function (S2), the Lorentz function (S3), and the LogNormal Distribution (S4):

$$f(x) = A \cdot \exp\left[-\frac{(x-x_0)^2}{2\sigma^2}\right], \tag{S2}$$

$$f(x) = \frac{2A}{\pi} \cdot \frac{\sigma}{4 \cdot (x-x_0)^2 + \sigma^2}, \tag{S3}$$

$$f(x) = \frac{A}{\sqrt{2\pi}\sigma x} \cdot \exp\frac{-\left[\ln\frac{x}{x_0}\right]^2}{2\sigma^2}, \tag{S4}$$

where $A$ is the normalization constant; $x_0$ is the mathematical expectation, which in our case corresponds to the average particle size, $x_0 = D_0^{TEM}$; $\sigma^2$ is the dispersion, which for an ensemble of particles means the size dispersion of particles.

During approximation of the experimental values for the number of particles by the (S2)–(S4) functions (see Fig. S1), it was found that the smallest error in the deviation of the experimental data from the dependences (S2)–(S4) was observed for the Lorentz function (S3) for $x = 0.05$ and the LogNormal Distribution (S4) for others. The criterion for the selection of the (S3) and (S4) functions was the highest approximation accuracy when the coefficient of determination $R^2$ takes the maximum value (see Table S4). All functions were normalized using the condition $\int_0^{+\infty} f(x)dx = 1$.

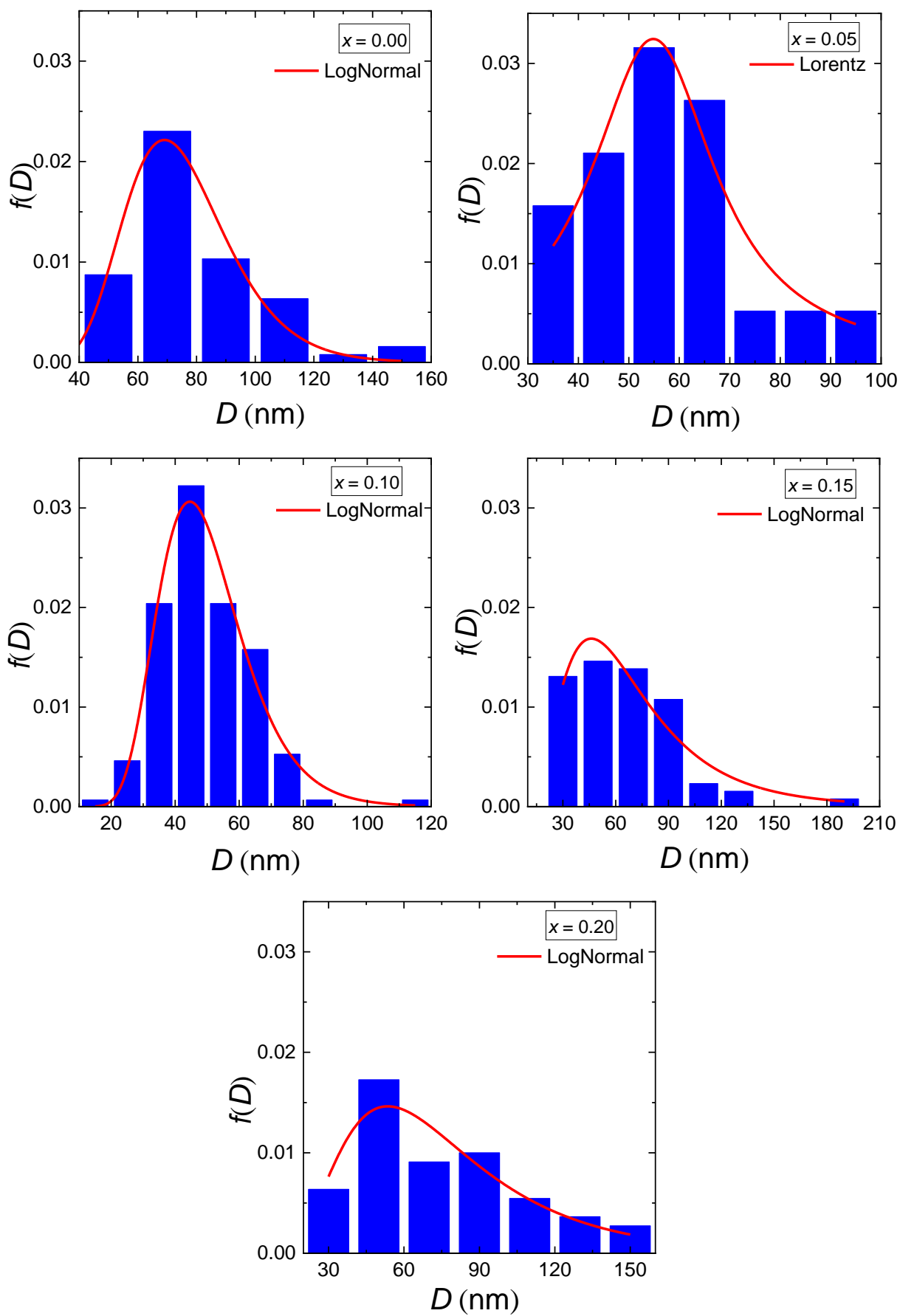

**Fig. S1**. Approximation of the experimental values $D$ in the $La_{0.8-x}\square_{x}Na_{0.2}Mn_{1+x}O_{3-\Delta}$ nanopowders with $x = 0.00$–$0.20$ by the Lorentz function (S3) and the LogNormal Distribution function (S4).

An average particle size $D_0^{TEM}$, dispersion $\sigma$ and determination coefficient $R^2$ for the La$_{0.8-x}\square_x$Na$_{0.2}$Mn$_{1+x}$O$_{3-\Delta}$ nanopowders with $x = 0.00$–$0.20$

| $x$ | Function | $D_0^{TEM}$ (nm) | $\sigma$ | $R^2$ |
|---|---|---|---|---|
| 0.00 | LogNormal | 73.3±1.9 | 24.5±2.2 | 0.96217 |
| 0.05 | Lorentz | 54.8±1.9 | 29.9±5.9 | 0.89084 |
| 0.10 | LogNormal | 48.3±0.8 | 28.3±1.6 | 0.97867 |
| 0.15 | LogNormal | 61.6±4.2 | 53.7±6.7 | 0.92612 |
| 0.20 | LogNormal | 69.2±6.4 | 50.7±8.1 | 0.82806 |

## S3. Determination of the valence state of lanthanum, sodium and manganese ions using XPS

Figure S2 shows La3d, Na1s and O1s X-ray photoelectron spectra for the $La_{0.8-x}Na_{0.2}Mn_{1+x}O_{3-\Delta}$ sample with $x = 0.00$. In the one-electron approximation, the X-ray photoelectron spectrum of La3d is $La3d_{5/2,\,3/2}$ spin doublet (see peaks A and C in Fig. S2a). In fact, the spectra have a much more complex profile due to the presence of charge transfer satellites (features B and D), plasmon lines and MNN Auger lines (features B′ and D′) [3, 4]. The thorough analysis of La3d spectra was performed for the $La_2O_3$ and $La(OH)_3$ earlier [3, 4]. The La3d spectra of the $La_{0.54}Bi_{0.17}Mn_{0.88}O_{3.40}$ single crystal and $La_2O_3$ powder were analyzed, and the satellite structure (B and D) was calculated [5].

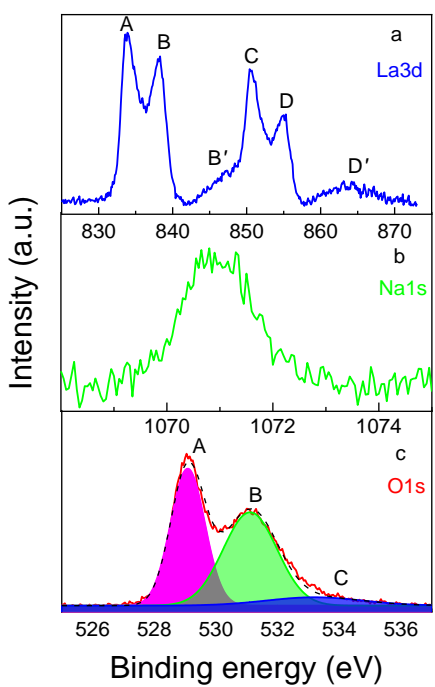

**Fig. S2**. The La3d (a), Na1s (b) and O1s (c) X-ray photoelectron spectra for the $La_{0.8-x}Na_{0.2}Mn_{1+x}O_{3-\Delta}$ sample with $x = 0.00$.

According to the work [4], the broad spectral features of B′ and D′ located at a distance of about 13.5 eV from the side of the higher binding energies of each main line of the A and B doublet originate from MNN Auger electron lines and plasmon excitations. Mullica *et.al.* [3] referred them exclusively to plasmon excitations. According to the paper [3], the sharp lines of B and D (see Fig. S2a) are charge transfer satellites reflecting the formation of shielded states $3d^{-1}4f^1L^{-1}$ (L = ligand) when a valence electron is shaken to a free 4f orbital after ionization of the La3d orbital. The La3d spectra were calculated taking into account charge transfer satellites and compared with La3d X-ray photoelectron spectra of the single-crystal $La_{0.54}Bi_{0.17}Mn_{0.88}O_{3.40}$ compound and $La_2O_3$ powder.

Although the La ion was trivalent in all considered compounds, the La3d spectrum of the $La_{0.54}Bi_{0.17}Mn_{0.88}O_{3.40}$ manganite is shifted by 1.9 eV, and the La3d spectrum in the $La_{0.8-x}Na_{0.2}Mn_{1+x}O_{3-\Delta}$ with $x = 0.00$ compound is shifted by 1.1 eV in the side of lower binding energies relative to the La3d spectrum of the $La_2O_3$ oxide ($E_b$(La3d) = 835.3 eV) [5]. This means that the local charge of lanthanum in the $La_2O_3$ is higher than in the La–Bi and La–Na manganites. This may be due to the different number of the nearest oxygen atoms around La in these compounds. According to the XRD data, the number of oxygen atoms in the first coordination sphere around La increases from 6 in $La_2O_3$ to 12 in both manganites. This leads to an increase in the electron density on La atoms and a shift of the XPS line towards lower binding energies. The relationship between the energy position of XPS lines and the number of the nearest neighboring atoms of the ligand in the case of organometallic nickel complexes was discussed in the work [6].

The energy distances between the main line and the peak of the charge transfer satellite $\Delta_{CTS}$ differ in all three spectra and are equal $\Delta_{CTS}$ = 4.3 eV for $La_{0.54}Bi_{0.17}Mn_{0.88}O_{3.40}$ [5], $\Delta_{CTS}$ = 4.4 eV for $La_{0.8-x}Na_{0.2}Mn_{1+x}O_{3-\Delta}$ ($x = 0.00$), and $\Delta_{CTS}$ = 3.3 eV for $La_2O_3$ [5]. This distance is approximately the energy required to excite the valence electron of the oxygen atom to the free La4f orbital. Obviously, this energy is less when the local positive charge on $La^{3+}$ is higher. Since the binding energy $E_{3d}$ of the main XPS lines increases with an increase in the local charge, such increase should be accompanied by a decrease in $\Delta_{CTS}$, what we can see.

Figure S2b shows the Na1s X-ray photoelectron spectrum, which is a weak single symmetric line. Figure S1c shows the O1s X-ray photoelectron spectrum consisting of three A, B and C components with energy positions of 529.6 eV, 531.4 eV and 533.1 eV, respectively. According to the paper [5], the A component refers to lattice oxygen embedded in the lattice of the $La_{0.8-x}Na_{0.2}Mn_{1+x}O_{3-\Delta}$ ($x = 0.00$) compound, B component is to be adsorbed oxygen and C component is to be hydroxyl groups or water vapor. The energetic positions of La3d, Na1s, Mn2p and O1s lines for the $La_{0.8-x}Na_{0.2}Mn_{1+x}O_{3-\Delta}$ nanopowders are presented in Table S5.

From the energy position of the $La3d_{3/2}$ electron line [7] and the calculation data of its profile [5], it can be concluded that lanthanum ions are in the trivalent state in all our studied compounds.

**Table S5**
Energy positions of the La3d, Mn2p, Na1s and O1s X-ray photoelectron lines in the $La_{0.8-x}Na_{0.2}Mn_{1+x}O_{3-\Delta}$ nanopowders

| x | La3d | | $Mn2p_{3/2}$ | Na1s | O1s | | |
| --- | --- | --- | --- | --- | --- | --- | --- |
| | $La3d_{3/2}$ | Sat | | | A | B | C |
| 0.00 | 834.2 | 838.6 | 642.5 | 1071.7 | 529.6 | 531.4 | 533.1 |
| 0.10 | 834.2 | 838.6 | 642.4 | 1071.5 | 529.7 | 531.3 | 532.6 |
| 0.20 | 834.0 | 838.3 | 642.2 | 1071.3 | 529.7 | 531.5 | 533.3 |

The determination of the valence state of transition metal ions is usually carried out by decomposing the $Me2p_{3/2}$ peak into symmetric components [8-10]. The components are classified as $Me^{2+}$, $Me^{3+}$, $Me^{4+}$ according to their energy positions, which increase in the order of $Me^{2+}$, $Me^{3+}$, $Me^{4+}$. This approach was used to determine the valence content of manganese in the *A* and *A'*-deficient $La_{0.67}Sr_{0.33}MnO_3$ compound [8]. In our opinion, this approximation is quite rough, since the Me2p X-ray photoelectron spectra of transition metals are multiplet consisting of a large number of components [11-13] and therefore having an asymmetric shape. The approximation of a large number of components by two or three components corresponding to several valence states of ion is quite arbitrary. In our previous works, an approach for fitting the experimental full profile of Fe2p and Mn2p X-ray photoelectron spectra has been proposed using either the calculated spectra (taking into account the multiplet) for each of the transition metal ions [11, 13] or the experimental $Me^{2+}$, $Me^{3+}$, $Me^{4+}$ (Me = Fe, Mn) spectra [5, 12].

In our case, assuming the presence of three manganese $Mn^{2+}$, $Mn^{3+}$, $Mn^{4+}$ ions in the $La_{0.8-x}\square_xNa_{0.2}Mn_{1+x}O_{3-\Delta}$ samples, the experimental Mn2p X-ray photoelectron spectrum, Spct(Mn), (see Fig. 2) was fitted by three experimental $Spct(Mn^{2+})$, $Spct(Mn^{3+})$ and $Spct(Mn^{4+})$ spectra of the $Mn_2V_2O_7$, $Mn_2O_3$ and $SrMnO_3$ samples, which contain $Mn^{2+}$, $Mn^{3+}$ and $Mn^{4+}$ manganese ions, respectively. Calculation of the multiplet structure of Mn2p X-ray photoelectron spectra of the $Mn^{2+}$, $Mn^{3+}$, $Mn^{4+}$ ions [11] showed good agreement between the theoretical and experimental profiles of the Mn2p spectra for the $Mn_2V_2O_7$, $Mn_2O_3$, $SrMnO_3$ samples, respectively.

$$\text{Spct}(Mn) = a \cdot \text{Spct}(Mn^{2+}) + b \cdot \text{Spct}(Mn^{3+}) + c \cdot \text{Spct}(Mn^{4+}), \tag{S5}$$

where the relative concentrations of $Mn^{2+}$, $Mn^{3+}$, $Mn^{4+}$ ions are equal to the coefficients $a$, $b$, $c$ and $a+b+c = 1$. The energy positions of the Mn2p$_{3/2}$ decomposition components of X-ray photoelectron spectra corresponding to the $Mn^{2+}$, $Mn^{3+}$, $Mn^{4+}$ ions are equal to 641.2 eV, 641.7 eV and 642.5 eV, which is in good agreement with the measured values of 641.2 eV, 641.7 eV and 642.0 eV [11], respectively. The fractions of $Mn^{2+}$, $Mn^{3+}$, $Mn^{4+}$ ions obtained by fitting the experimental Mn2p spectra for the $La_{0.8-x}\square_x Na_{0.2}Mn_{1+x}O_{3-\Delta}$ samples are presented in Table S6.

**Table S6**
Fractions of the $Mn^{2+}$, $Mn^{3+}$ and $Mn^{4+}$ ions according to the results of the decomposition of Mn2p X-ray photoelectron spectra for the $La_{0.8-x}Na_{0.2}Mn_{1+x}O_{3-\Delta}$ nanopowders

| $x$ | $Mn^{2+}$ | $Mn^{3+}$ | $Mn^{4+}$ |
|---|---|---|---|
| 0.00 | 0.297 | 0.390 | 0.313 |
| 0.10 | 0.273 | 0.371 | 0.356 |
| 0.20 | 0.244 | 0.387 | 0.369 |

According to Table S6 and different intensities of the analytical O1s line, the chemical composition taking into account the electroneutrality of the $La_{0.8-x}\square_x Na_{0.2}Mn_{1+x}O_{3-\Delta}$ samples can be determined as it has been shown in the work [5]. The intensity of the O1s line changes from the intensity value of the entire experimental spectrum to its value when *B*-component is completely subtracted from the spectrum of the O1s line (see Fig. S2c). The *B*-component can refer to both the oxygen adsorbed on the surface and the anion vacancies in the sample [11]. For each new value of the O1s line intensity, the charge balance (the sum of the charges of all ions in the structural unit of the sample) and oxygen content, the values of oxygen non-stoichiometry $\Delta$ are calculated (see Fig. 2b-d). For example, the elemental composition for the $La_{0.8}Na_{0.2}MnO_3$ sample (at point 2 in Fig. 2b) calculated using the total intensity of O1s spectrum yields the $La_{0.62}Na_{0.17}Mn_{0.69}O_{3.52}$. If the oxygen content is stoichiometric (point 3 in Fig. 2b), the formula is $La_{0.84}Na_{0.22}Mn_{0.93}O_{3.00}$. Finally, if the intensity of only *A* component of the O1s spectrum is used (point 1 in Fig. 2b) coming from the in-lattice oxygen, we can get $La_{1.03}Na_{0.27}Mn_{1.14}O_{2.56}$.

Additionally, EDS data for all $La_{0.8-x}\square_{x}Na_{0.2}Mn_{1+x}O_{3-\Delta}$ compositions are presented in Table S7 that confirms their stoichiometry.

**Table S7**
Chemical composition (at. %) according to EDS data in the $La_{0.8-x}\square_{x}Na_{0.2}Mn_{1+x}O_{3-\Delta}$ nanopowder for non-defect and real defect cases

| $x$ | Case | Element (at. %) | | | |
|---|---|---|---|---|---|
| | | La | Na | Mn | O |
| 0.00 | Non-defect | 16.00 | 4.00 | 20.00 | 60.00 |
| | Defect | 15.26±0.95 | 4.39±0.78 | 20.18±0.42 | 60.17±1.35 |
| 0.05 | Non-defect | 15.00 | 4.00 | 21.00 | 60.00 |
| | Defect | 14.17±0.89 | 4.08±0.69 | 20.90±0.68 | 60.86±1.37 |
| 0.10 | Non-defect | 14.00 | 4.00 | 22.00 | 60.00 |
| | Defect | 13.17±0.64 | 4.23±0.64 | 22.81±0.56 | 59.79±0.77 |
| 0.15 | Non-defect | 13.00 | 4.00 | 23.00 | 60.00 |
| | Defect | 12.74±0.57 | 4.43±0.51 | 23.09±0.26 | 59.99±1.85 |
| 0.20 | Non-defect | 12.00 | 4.00 | 24.00 | 60.00 |
| | Defect | 11.50±0.41 | 4.81±0.47 | 24.52±0.63 | 59.17±1.21 |

**Table S8**
Elemental composition and concentration of ions (at. %) depending on the approaches for the $La_{0.8-x}\square_{x}Na_{0.2}Mn_{1+x}O_{3-\Delta}$

| $x$ | Approach | Elemental composition | Concentration of ions (at. %) | | | |
|---|---|---|---|---|---|---|
| | | | La | Na | Mn | O |
| 0.00 | Non-defect | $\{La^{3+}_{0.80}Na^{1+}_{0.20}\}_A[Mn^{3+}_{0.60}Mn^{4+}_{0.40}]_B O^{2-}_{3.00}$ | 16.00 | 4.00 | 20.00 | 60.00 |
| | IT | $\{La^{3+}_{0.80}Na^{1+}_{0.20}\}_A[Mn^{3+}_{0.68}Mn^{4+}_{0.32}]_B O^{2-}_{2.96}V^{(a)}_{0.04}$ | 16.00 | 4.00 | 20.00 | 59.20 |
| | EDS | $La^{3+}_{0.763}Na^{1+}_{0.219}(Mn^{3+}_{0.544}Mn^{4+}_{0.465})_{1.009}O^{2-}_{3.01}$ | 15.26 | 4.39 | 20.18 | 60.17 |
| | XPS | $La^{3+}_{0.882}Na^{1+}_{0.234}(Mn^{2+}_{0.289}Mn^{3+}_{0.380}Mn^{4+}_{0.305})_{0.974}O^{2-}_{2.91}V^{(a)}_{0.09}$ | 17.64 | 4.68 | 19.48 | 58.20 |
| 0.10 | Non-defect | $\{La^{3+}_{0.70}Na^{1+}_{0.20}Mn^{2+}_{0.10}\}_A[Mn^{3+}_{0.50}Mn^{4+}_{0.50}]_B O^{2-}_{3.00}$ | 14.00 | 4.00 | 22.00 | 60.00 |
| | IT | $\{La^{3+}_{0.70}Na^{1+}_{0.20}Mn^{2+}_{0.10}\}_A[Mn^{3+}_{0.60}Mn^{4+}_{0.40}]_B O^{2-}_{2.95}V^{(a)}_{0.05}$ | 14.00 | 4.00 | 22.00 | 59.00 |
| | EDS | $La^{3+}_{0.659}Na^{1+}_{0.212}(Mn^{2+}_{0.101}Mn^{3+}_{0.571}Mn^{4+}_{0.469})_{1.141}O^{2-}_{2.99}V^{(a)}_{0.01}$ | 13.17 | 4.23 | 22.81 | 59.79 |
| | XPS | $La^{3+}_{0.694}Na^{1+}_{0.278}(Mn^{2+}_{0.305}Mn^{3+}_{0.415}Mn^{4+}_{0.398})_{1.118}O^{2-}_{2.91}V^{(a)}_{0.09}$ | 13.88 | 5.56 | 22.36 | 58.20 |
| 0.20 | Non-defect | $\{La^{3+}_{0.60}Na^{1+}_{0.20}Mn^{2+}_{0.20}\}_A[Mn^{3+}_{0.40}Mn^{4+}_{0.60}]_B O^{2-}_{3.00}$ | 12.00 | 4.00 | 24.00 | 60.00 |
| | IT | $\{La^{3+}_{0.60}Na^{1+}_{0.20}Mn^{2+}_{0.20}\}_A[Mn^{3+}_{0.54}Mn^{4+}_{0.46}]_B O^{2-}_{2.93}V^{(a)}_{0.07}$ | 12.00 | 4.00 | 24.00 | 58.60 |
| | EDS | $La^{3+}_{0.575}Na^{1+}_{0.241}(Mn^{2+}_{0.201}Mn^{3+}_{0.568}Mn^{4+}_{0.457})_{1.226}O^{2-}_{2.95}V^{(a)}_{0.05}$ | 11.50 | 4.81 | 24.52 | 59.17 |
| | XPS | $La^{3+}_{0.648}Na^{1+}_{0.251}(Mn^{2+}_{0.286}Mn^{3+}_{0.453}Mn^{4+}_{0.432})_{1.171}O^{2-}_{2.93}V^{(a)}_{0.07}$ | 12.96 | 5.02 | 23.42 | 58.60 |

## S4. Magnetic properties of the La$_{0.8-x}$□$_x$Na$_{0.2}$Mn$_{1+x}$O$_{3-\Delta}$ nanopowders

The field dependences of magnetization $M(H)$ for the La$_{0.8-x}$□$_x$Na$_{0.2}$Mn$_{1+x}$O$_{3-\Delta}$ nanopowder depending on the $x = 0.00$–$0.20$ are presented in Fig. S3. With increase in $x$, a non-monotonic change in the saturation magnetization $M_S$ (see upper inset of Fig. S3e), as well as the coercivity $H_C$ and the remanent magnetization $M_r$ demonstrating magnetically soft nature of all La$_{0.8-x}$□$_x$Na$_{0.2}$Mn$_{1+x}$O$_{3-\Delta}$ (see bottom inset of Fig. S3e) is observed. Such non-monotonic change of the $M_S$ values at $T = 2$ K in the non-stoichiometric La$_{0.8-x}$□$_x$Na$_{0.2}$Mn$_{1+x}$O$_{3-\Delta}$ compound can be associated with different competing types of 180° quasi-stable FM Mn$^{3+}$↔O$^{2-}$↔Mn$^{4+}$, strong AFM Mn$^{2+}$↔O$^{2-}$↔Mn$^{2+}$, quasi-stable AFM Mn$^{4+}$↔O$^{2-}$↔Mn$^{2+}$, weak AFM Mn$^{4+}$↔O$^{2-}$↔Mn$^{4+}$ and 90° AFM Mn$^{2+}$↔O$^{2-}$↔Mn$^{4+}$ interactions [14-16], which strongly depend on changing structural parameters. Indeed, at small level of $x \leq 0.05$ (see Table S1), the bond distance <$d_{Mn-O}$> ($Mn_B^{3+/4+} - O^{2-}$) decreases that enhances the FM DE Mn$^{3+}$↔O$^{2-}$↔Mn$^{4+}$ and increases the $M_S$. At higher $x \geq 0.10$, the bond distance <$d_{Mn-O}$> ($Mn_B^{3+/4+} - O^{2-}$) and angle Mn−O−Mn ($Mn_B^{3+/4+} - O^{2-} - Mn_B^{3+/4+}$) increase and decrease, respectively, that weakens the DE and reduces the $M_S$. Moreover, the multiple DE interactions via the Mn$^{3+}$↔O$^{2-}$↔Mn$^{2+}$↔O$^{2-}$↔Mn$^{4+}$ path can also lead to the increase in ferromagnetism and conductivity because of electronic band features of Mn$^{2+}$ ions, which are located at the 12-coordinated cubooctahedral La-deficient sites of the perovskite structure with a reversed $t_{2g}$–$e_g$ orbital and sufficiently decreased splitting ~ 20 % of 10$Dq_{oct}$ compared to the Mn$^{3+/4+}$ ions located at the 8-coordinated octahedral $B$-sites [8, 17-20]. At the same time, the bond distance <$d_{Mn-O}$> ($La_A^{3+}/Na_A^+/Mn_A^{2+} - O^{2-}$) and angle Mn−O−Mn ($La_A^{3+}/Na_A^+/Mn_A^{2+} - O^{2-} - Mn_B^{3+/4+}$) decrease and increase, respectively, which should favorable influence the multiple DE. It should be noted that as it was shown in the works [19, 20], the conventional DE Mn$^{3+}$↔O$^{2-}$↔Mn$^{4+}$ dominates in the low temperature FM state and the additional multiple DE Mn$^{3+}$↔O$^{2-}$↔Mn$^{2+}$↔O$^{2-}$↔Mn$^{4+}$ is the main mechanism in the high temperature region near $T_C$ for manganites with overstoichiometric manganese.

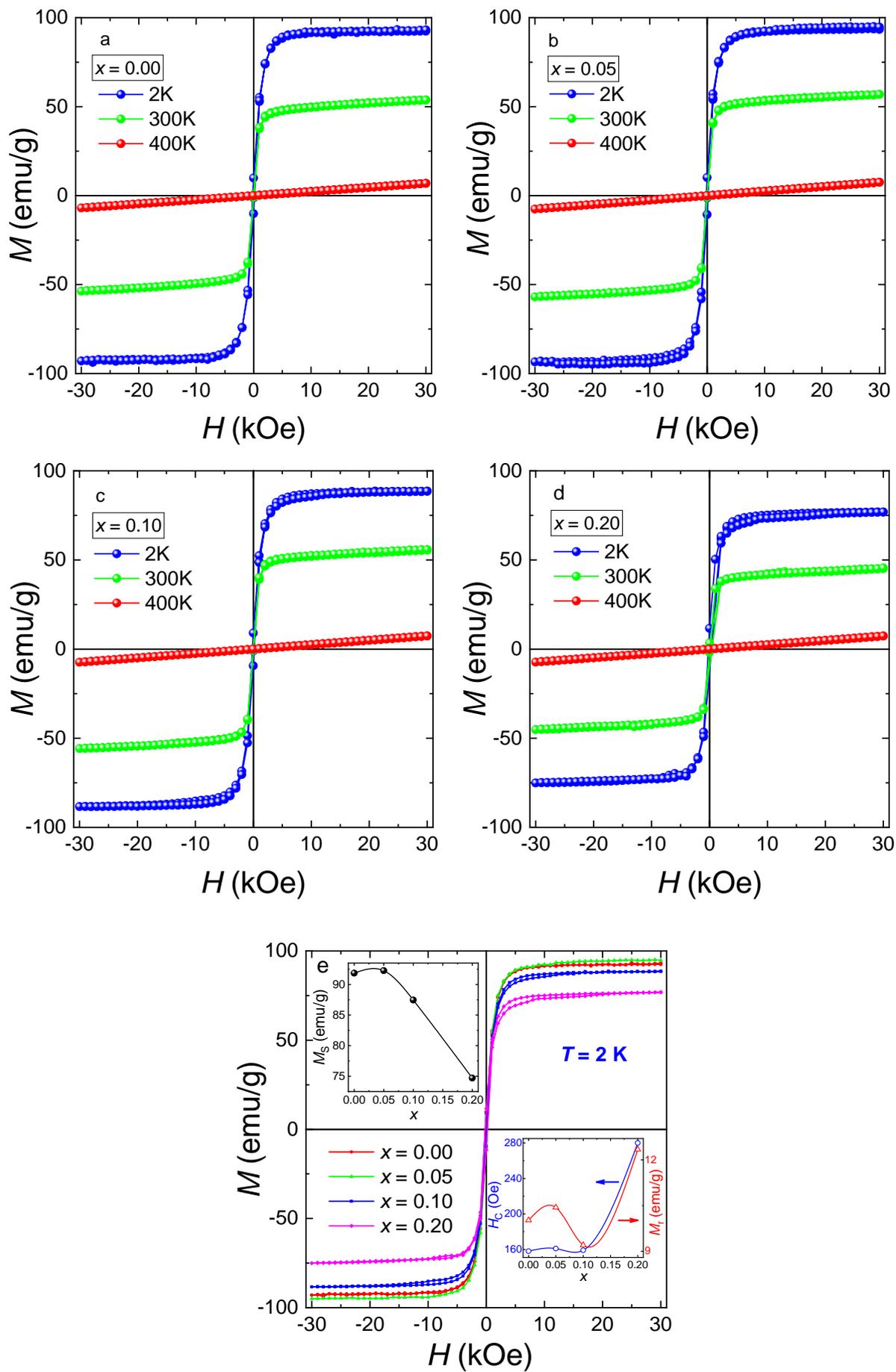

**Fig. S3**. Field dependences of magnetization $M(H)$ at 2, 300 and 400 K for the La$_{0.8-x}\square_x$Na$_{0.2}$Mn$_{1+x}$O$_{3-\Delta}$ nanopowders with $x = 0.00$ (a), 0.05 (b), 0.10 (c) and 0.20 (d), as well as the comparable $M(H)$ at 2 K for all compositions (e) (the upper and bottom insets show the concentration dependences of the saturation magnetization $M_S$, coercivity $H_C$ and remanent magnetization $M_r$, respectively).

**Table S9**

Saturation magnetization $M_S$, remanent magnetization $M_r$ and coercivity $H_C$ at different temperatures for the La$_{0.8-x}\square_x$Na$_{0.2}$Mn$_{1+x}$O$_{3-\Delta}$ nanopowders with $x = 0.00$–0.20

| $x$ | $M_S$(2K) (emu/g) | $M_r$(2K) (emu/g) | $M_r$(300K) (emu/g) | $H_C$(2K) (kOe) | $H_C$(300K) (kOe) |
|---|---|---|---|---|---|
| 0.00 | 91.86 | 10.017 | 0.690 | 0.158 | 0.018 |
| 0.05 | 92.26 | 10.439 | 0.440 | 0.161 | 0.011 |
| 0.10 | 87.46 | 9.204 | 0.233 | 0.159 | 0.006 |
| 0.20 | 74.71 | 12.340 | 5.905 | 0.280 | 0.234 |

**Table S10**

The magnetic moments $\mu$ experimentally and theoretically defined for the real defect La$_{0.8-x}\square_x$Na$_{0.2}$Mn$_{1+x}$O$_{3-\Delta}$ compounds with $\Delta \neq 0$ depending on $x = 0.00$–0.20 in the different temperature regions and magnetic states

| $x$ | $T_C$ (K) | $\theta$ (K) | $\mu_{FM}^{exp}$ ($\mu_B$) | $\mu_{FM}^{theor}$ ($\mu_B$) | $\mu_{PM}^{exp}$ ($\mu_B$) | $\mu_{PM}^{theor}$ ($\mu_B$) |
|---|---|---|---|---|---|---|
| 0.00 | 327 | 331 | 3.59 | 3.68 | 5.23 | 4.60 |
| 0.05 | 331 | 336 | 3.53 | 3.70 | 5.09 | 4.61 |
| 0.10* | 331 | 334 | 3.35 | 3.73 | 5.01 | 4.64 |
| 0.20* | 328 | 332 | 2.86 | 3.78 | 4.03 | 4.70 |

*taking into account content of Mn$_3$O$_4$ second phase

The temperature dependences of magnetization $M(T)$ for all La$_{0.8-x}\square_x$Na$_{0.2}$Mn$_{1+x}$O$_{3-\Delta}$ ($x = 0.00$–0.20) compositions measured in the zero-field-cooled (ZFC) and field-cooled (FC) modes under a magnetic field of $H = 50$ Oe are presented in Fig. S4. A significant thermo-magnetic irreversibility between ZFC and FC curves is observed for all compositions that may be caused by the magnetic frustration in them [21]. With increase in $x$ from 0.00 to 0.20, a non-monotonic change of FM phase fraction is also observed. Additionally, a small anomaly peak on FC curves becomes clearer at ~ 50 K with increase in $x$. It can be associated with the several factors, such as: (i) the presence of Mn$^{2+}$ ions forming 90° AFM planar nanostructured clusters inside the perovskite structure [22]; (ii) Mn$^{2+}$ ions located on the surface of all samples (see XPS data in Table S8) [18]; and (iii) a nucleation of ferrimagnetic Mn$_3$O$_4$ second-phase (see Table S1) which is in PM state at $T > 50$ K and does not influence on the magnetization in a high temperature region near $T_C$ [23-25].

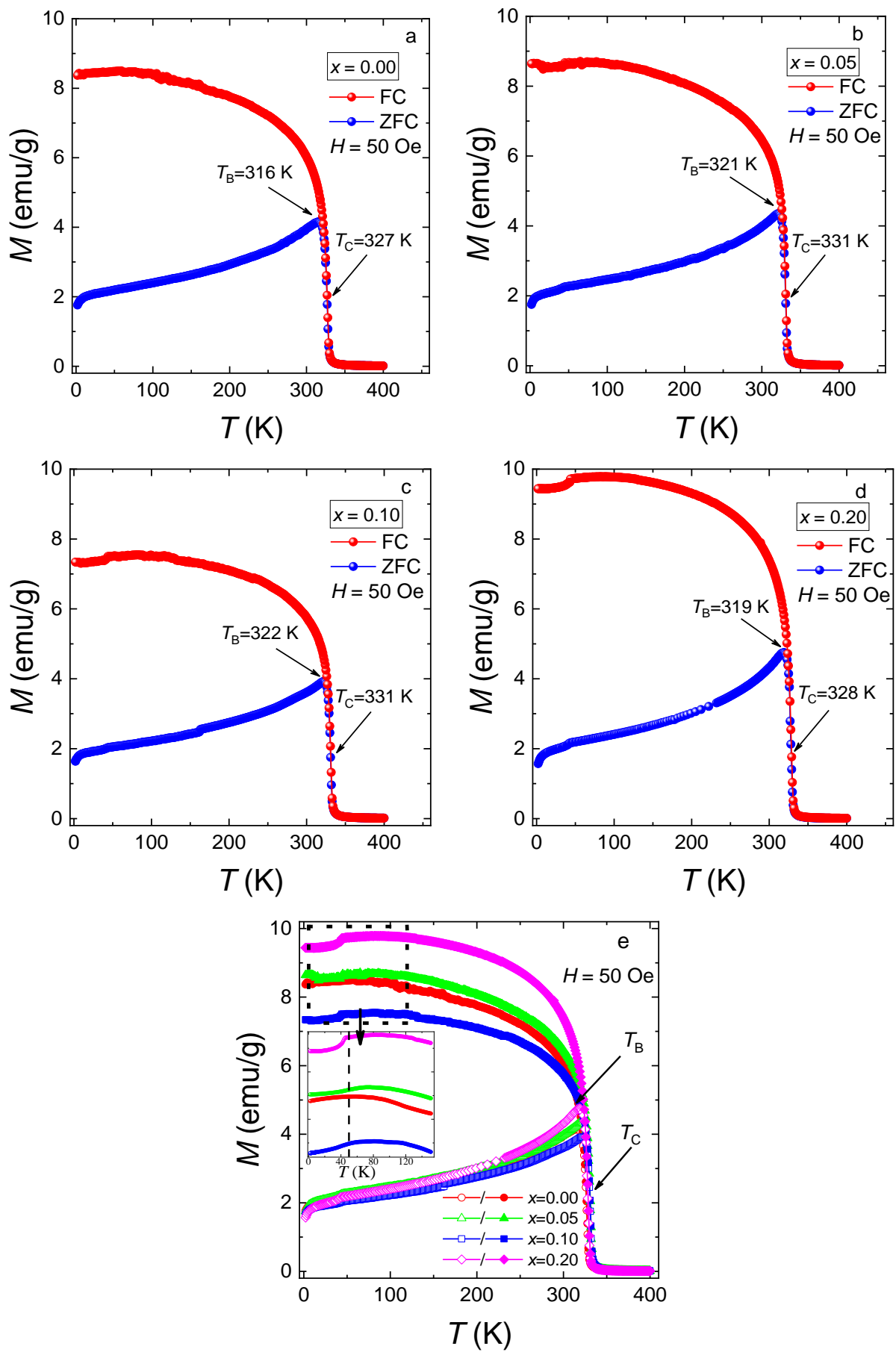

**Fig. S4.** Temperature dependences of $M_{FC}(T)$ and $M_{ZFC}(T)$ in the field of $H = 50$ Oe for the La$_{0.8-x}\square_x$Na$_{0.2}$Mn$_{1+x}$O$_{3-\Delta}$ nanopowders with $x = 0.00$ (a), 0.05 (b), 0.10 (c) and 0.20 (d), as well as the comparable $M_{FC}(T)$ and $M_{ZFC}(T)$ for all compositions (e) (the insert shows the enlarged area of $M(T)$ where a small peak near 50 K is observed (dashed line)).

The maximum temperature $T_C$ and blocking temperature $T_B$ were determined using the $M(T)$ dependences as $|dM/dT|$=max and the maximum value of the ZFC curve [23], respectively. With increase in $x$, the both temperatures of $T_C$ and $T_B$ increase first from 327 and 316 K ($x = 0.00$) to 331 and 322 K ($x = 0.10$) and then decreased to 328 and 319 K ($x = 0.20$). At $T < T_B$, the nanoparticles are in the SPM blocked state. In the range of $T_B < T < T_C$, the SPM nanoparticles are magnetized equilibriumly [23].

In the PM and high temperature ($T > T_C$) region, the effective magnetic moment of manganese $\mu_{PM}(Mn)$ can be obtained from the equation [23, 26]:

$$\mu_{PM}(Mn) = \sqrt{\frac{x_{Mn^{2+}}}{x} \cdot \mu^2_{Mn^{2+}} + \frac{x_{Mn^{3+}}}{x} \cdot \mu^2_{Mn^{3+}} + \frac{x_{Mn^{4+}}}{x} \cdot \mu^2_{Mn^{4+}}}, \quad (S6)$$

where $x$ is the concentration of overstoichiometric manganese; $x_{Mn}^{2+}$, $x_{Mn}^{3+}$ and $x_{Mn}^{4+}$ are the fractions of manganese, respectively. In the ground state, the effective magnetic moments of manganese are equal to $\mu_{Mn}^{2+} = 5.748\mu_B$, $\mu_{Mn}^{3+} = 4.90\mu_B$ and $\mu_{Mn}^{4+} = 3.87\mu_B$ [27]. Using the Curie-Weiss law $\chi = C / (T - \theta)$ during an approximation of experimental data $\chi^{-1}(T)$ (see Fig. S5), the Curie constant $C$, the PM Curie temperature $\theta$ and the effective magnetic moment of manganese $\mu_{PM}^{exp}$ with high accuracy were determined (see Fig. S5 and Table S10). For defining the $\mu_{PM}^{theor}$, we used the electroneutrality principle ($3 \cdot x_{La}^{3+} + 1 \cdot x_{Na}^{+} + 2 \cdot x_{Mn}^{2+} + 3 \cdot x_{Mn}^{3+} + 4 \cdot x_{Mn}^{4+} - 2 \cdot (3-\Delta) = 0$, where $x_{La}^{3+}$, $x_{Na}^{+}$, $x_{Mn}^{2+}$, $x_{Mn}^{3+}$, $x_{Mn}^{4+}$ are the concentrations of lanthanum, sodium and manganese ions, respectively) and then Eq. (S6) for the real defect $La_{0.8-x}\square_x Na_{0.2}Mn_{1+x}O_{3-\Delta}$ compounds ($x = 0.00$–$0.20$) with $\Delta \neq 0$.

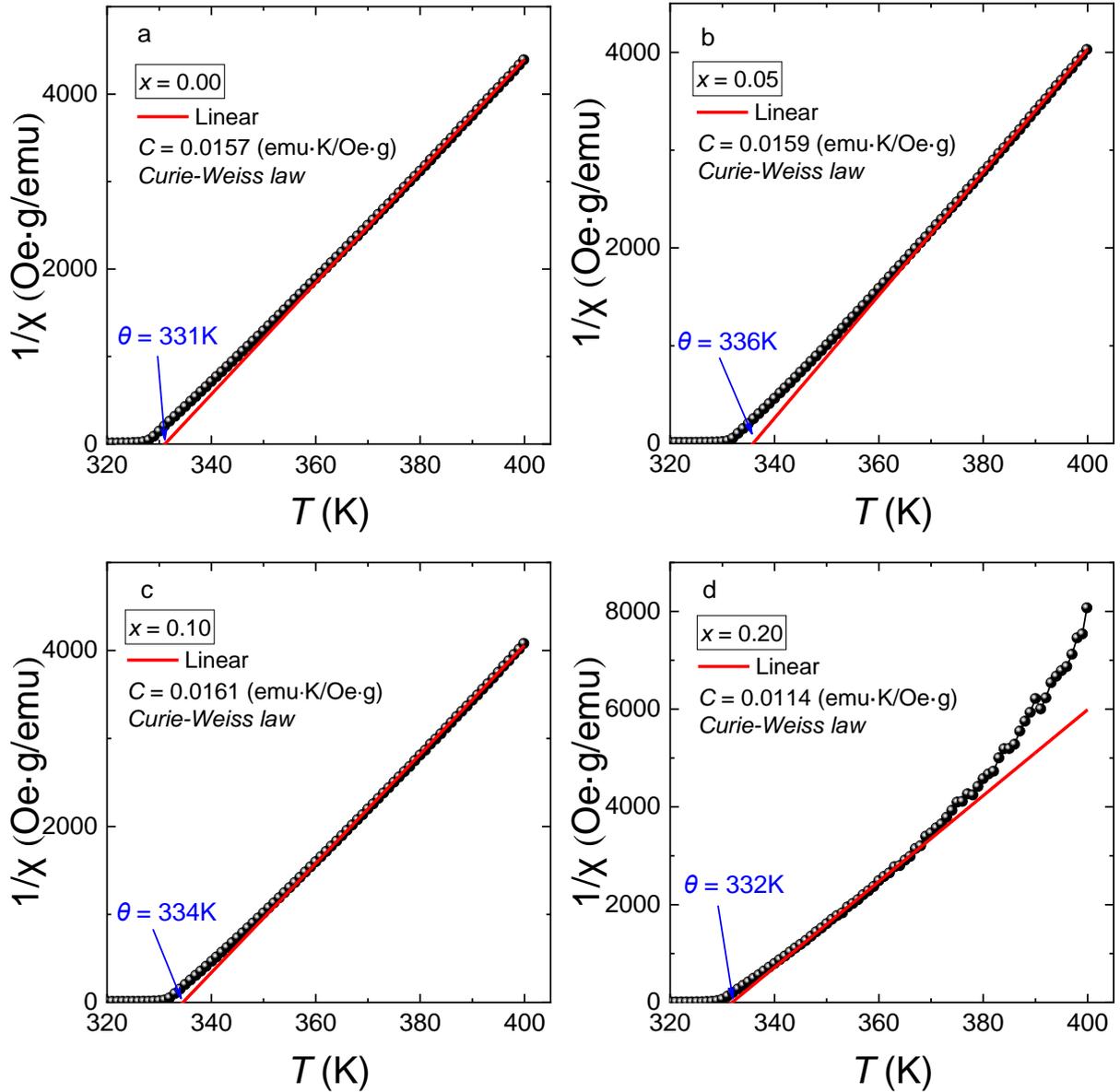

**Fig. S5**. An approximation of the inverse magnetic susceptibility $\chi^{-1}(T) = H/M(T)$ by the Curie-Weiss law ($\theta$ is the PM Curie temperature and $C$ is the Curie constant) for the $La_{0.8-x}\square_{x}Na_{0.2}Mn_{1+x}O_{3-\Delta}$ nanopowders with $x = 0.00$ (a), 0.05 (b), 0.10 (c) and 0.20 (d).

For defining the contribution from the surface layer containing $Mn^{2+}$ ions in the $\mu_{PM}^{exp}$, the XPS and IT data (see Table S8) were used. Since the XPS beam penetration $\sim 3\lambda$ [28], where the $\lambda$ is the free path length of electrons depending on the energy of the electronic line, the surface layer thickness for Mn ions is $d \sim 5.8$ nm. For the spherical-like nanoparticle with a size of $D = 61$ nm (see Table S1 for $x = 0.00$), the contribution from the surface layer in $\mu_{PM}^{exp}$ can be calculated as:

$$\frac{[D^3-(D-d)^3]\cdot x_{Mn^{2+}}\cdot \mu_{Mn^{2+}}}{[D^3-(D-d)^3]\cdot (x_{Mn^{2+}}\cdot \mu_{Mn^{2+}} + x_{Mn^{3+}}\cdot \mu_{Mn^{3+}} + x_{Mn^{4+}}\cdot \mu_{Mn^{4+}}) + [D^3-d^3]\cdot (x_{Mn^{3+}}\cdot \mu_{Mn^{3+}} + x_{Mn^{4+}}\cdot \mu_{Mn^{4+}})},$$

where $D$ is the size of spherical-like nanoparticle; $d$ is the surface layer thickness; $x_{Mn}^{2+}$, $x_{Mn}^{3+}$ and $x_{Mn^{4+}}$ are the fractions of manganese per unit cell, respectively; the effective magnetic moments $\mu_{Mn^{2+}}$ = 5.748$\mu_B$, $\mu_{Mn^{3+}}$ = 4.90$\mu_B$ and $\mu_{Mn^{4+}}$ = 3.87$\mu_B$. This gives about 9% = 0.537$\mu_B$ and consequently 4.693$\mu_B$ for the core containing only $Mn^{3+/4+}$ ions.

## S5. The spin-dependent exchange in manganites

The Hamiltonian of magnetic subsystem can be written as

$$\hat{H} = \sum_n \hat{H}_{0n} + \sum_{n,m} J_{nm}(\mathbf{r}_n, S_n, \mathbf{r}_m, S_m) \hat{\mathbf{s}}_n \hat{\mathbf{s}}_m, \quad (S7)$$

where $n$, $m$ are the numbers of ions; $\hat{H}_{0n}$ is the operator of single-ion electronic states, the own states and energies of which are determined by the spin $S_n$ of ion; $\hat{\mathbf{s}}_n$ is the spin operator of ion; $J_{nm}(\mathbf{r}_n, S_n, \mathbf{r}_m, S_m)$ is the exchange parameter between ions which depends on the integrals of overlapping wave functions, i.e. the the positions of ions $\mathbf{r}_n$, $\mathbf{r}_m$ and the values of the spins $S_n$, $S_m$ for each pair of ions. For ferromagnetically interacting pairs of ions, the exchange parameter is negative, and for antiferromagnetically interacting pairs of ions the parameter value is positive.

The form of the wave function of the ground state of an ion is determined by the single-ion part of Hamiltonian Eq. (S7) and depends on the spin $S$ of the ion. When the value of the spin $S$ changes, the both distribution of the electron density and the value of the overlap integral of the wave functions are changed. Thus, a change in the spin will affect the exchange between ions. The magnetic and electronic subsystems of ions are interconnected.

The energy of the magnetic subsystem per unit cell is equal to

$$E = \sum_i E_{0i}(S_i) + \sum_{ij} J_{ij}(S_i, S_j) <\mathbf{s}_i><\mathbf{s}_j>, \quad (S8)$$

where the summation is on the pairs of ions in the cell; $E_{0i}(S_i)$ is the energy of the main electronic state with the spin $S_i$; $<\mathbf{s}_i>$ is the vector of the average spin of the $i$-th ion; $J_{ij}(S_i, S_j)$ is the effective exchange constant for pairs of ions with $S_i$ and $S_j$ spins.

In the experiment and theory, the average magnetization per one ion is determined (see Table S8):

$$\boldsymbol{\mu} = g\mu_B <\mathbf{s}>, \quad (S9)$$

where $<\mathbf{s}> = \frac{1}{k}\sum_{i=1}^{k}<\mathbf{s}_i>$ is the average spin of Mn; $k$ is the number of ions in the cell, $g$ is the factor.

Using Eq. (S9), the energy of magnetic subsystem per unit cell can be written as:

$$E = \sum_i E_{0i}(S_i) - kJ_{eff} <\mathbf{s}>^2, \tag{S10}$$

where $J_{eff}$ is the effective exchange constant, which corresponds to the ferromagnetic and antiferromagnetic type of magnetic ordering for $J_{eff} < 0$ and $J_{eff} > 0$, respectively.

According to (S8) and (S10), it follows that the value of the effective ferromagnetic exchange constant satisfies

$$kJ_{eff} <\mathbf{s}>^2 = -\sum_{ij} J_{ij}(S_i, S_j) <\mathbf{s}_i><\mathbf{s}_j>. \tag{S11}$$

From Eq. (S11) can be seen that the effective exchange constant depends on the spins of the electronic state of ions. The magnitude of the exchange field $\mathbf{h}_{exch} = -g\mu_B \partial E/\partial <\mathbf{s}>$ also depends on the spin values of ions.

In the paramagnetic phase $<\mathbf{s}> = 0$, and the energy from Eq. (S10) is determined only by the first term and depends only on the energy of single-ion states. In the ferromagnetic phase $<\mathbf{s}> \neq 0$, the magnetic and spin states should be considered consistently.

It was found experimentally for the samples under study that the spins in the ferromagnetic phase are smaller than in the paramagnetic phase. The gain in energy for the low-spin state can be provided by the first term in Eq. (S10), as it happens in the case of spin phase transitions with the ground low-spin state of ions at low temperatures [29]. It can also be assumed that the exchange between ions increases in the samples in the low-spin state.

According to Eqs. (S8) and (S11) that the samples are non-Heisenberg magnets. The reason for their strong nonlinearity is the need to match the ferromagnetic state of the particle with the spin states of its magnetic ions. This nonlinearity can be responsible for a decrease in the spin in the ground state of ions at low temperatures and it can lead to non-Heisenberg behavior of the studied manganites near the Curie point.

## S6. Critical behavior of magnetization for the La$_{0.8-x}$□$_x$Na$_{0.2}$Mn$_{1+x}$O$_{3-\Delta}$ nanopowders

The field dependences of the magnetization $M(H)$ and ln$M$(ln$H$) for determining critical index of δ in the La$_{0.8-x}$□$_x$Na$_{0.2}$Mn$_{1+x}$O$_{3-\Delta}$ nanopowders with $x$ = 0.00–0.20 are presented in Fig. S6.

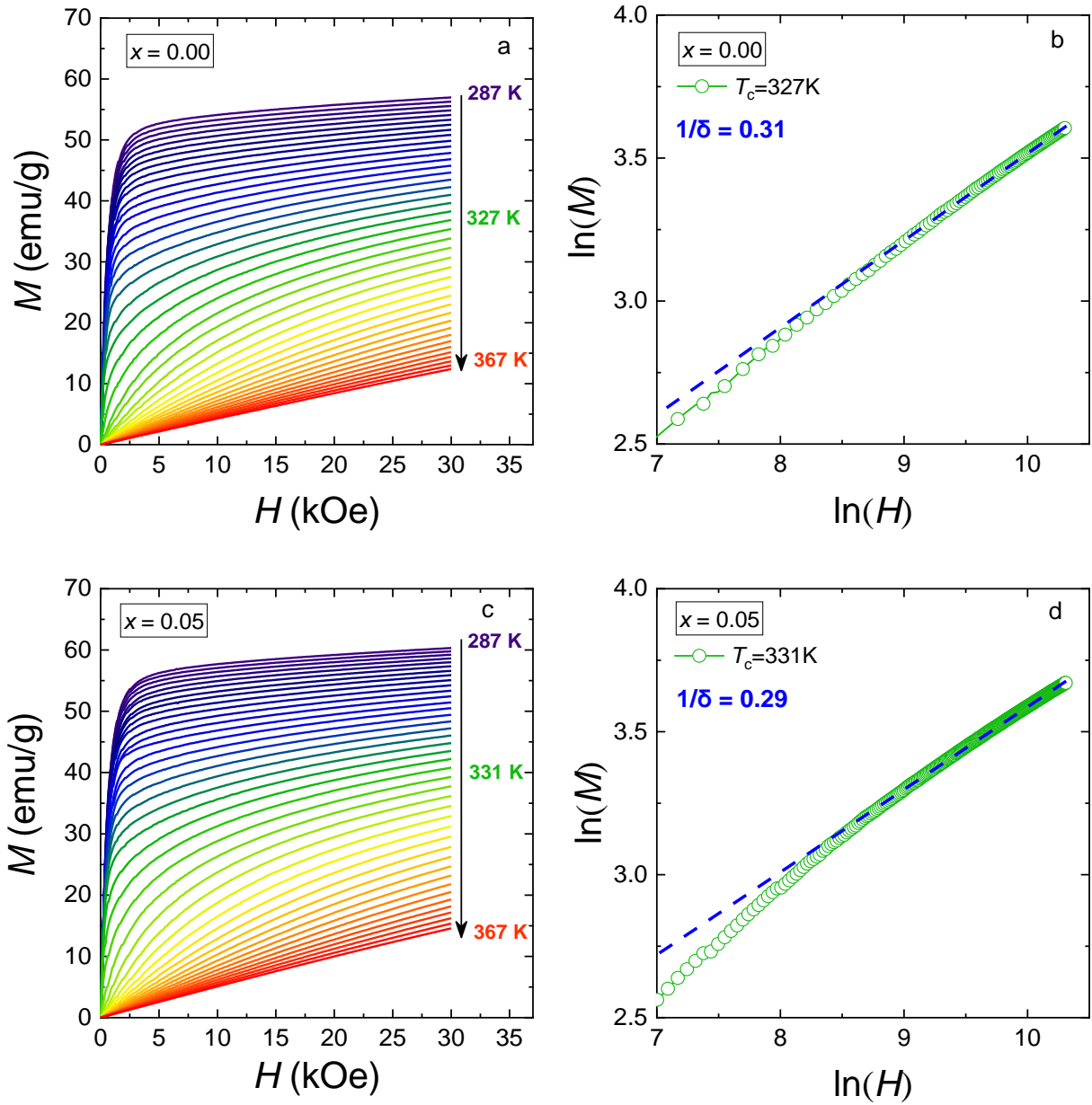

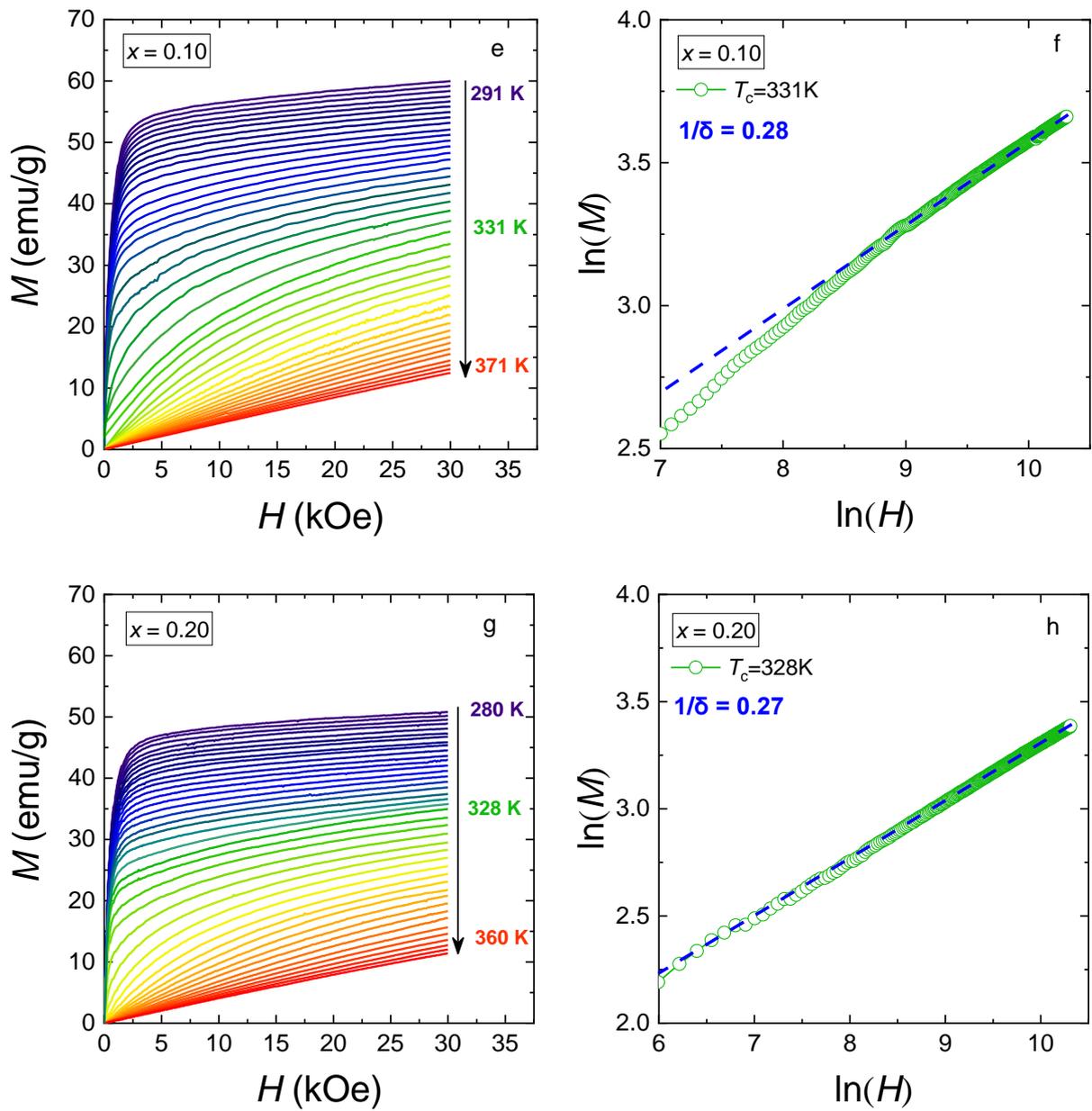

**Fig. S6.** Field dependences of the magnetization $M(H)$ and $\ln M(\ln H)$ for determining critical index of $\delta$ (green circles are experimental data and blue dashed line is their asymptote) in the La$_{0.8-x}\square_x$Na$_{0.2}$Mn$_{1+x}$O$_{3-\Delta}$ nanopowders with $x$ = 0.00 (a) and (b), 0.05 (c) and (d), 0.10 (e) and (f), 0.20 (g) and (h), respectively.

**Table S11**

Critical temperatures and parameters for the $La_{0.8-x}\square_x Na_{0.2} Mn_{1+x} O_{3-\Delta}$: Curie temperature $T_C$ ($|dM/dT|$=max); blocking temperature $T_B$ (peak in ZFC-curve); critical temperature $T_0$ ($M_S(T)\rightarrow 0$); Mean field model temperature $T_{MFM}$ ($\beta = 0.5$, $\gamma = 1.0$); Tricritical mean field model temperature $T_{TMFM}$ ($\beta = 0.25$, $\gamma = 1.0$); Ising temperature $T_{Ising}$ ($\beta = 0.325$, $\gamma = 1.24$); Heisenberg temperature $T_{Heisenberg}$ ($\beta = 0.365$, $\gamma = 1.336$); Modified Arrott plot model temperature $T_{MAP}$ ($\delta$, $\beta$, $\gamma$) and MCE temperature $T_{MCE}$ ($|-\Delta S_M|$ = max)

| x | $T_C$ (K) | $T_B$ (K) | $T_0$ (K) | $T_{MFM}$ (K) | $T_{TMFM}$ (K) | $T_{Ising}$ (K) | $T_{Heisenberg}$ (K) | MAP | | | | $T_{MCE}$ (K) | |
|---|---|---|---|---|---|---|---|---|---|---|---|---|---|
| | | | | | | | | $T_{MAP}$(K) | $\delta$ | $\beta$ | $\gamma$ | $H$=2kOe | $H$=30kOe |
| 0.00 | 327 | 316 | 328 | 329 | 321 | 323 | 323 | 327 | 3.20 | 0.37 | 0.81 | 325 | 329 |
| 0.05 | 331 | 321 | 330 | 333 | 327 | 327 | 329 | 331 | 3.46 | 0.32 | 0.78 | 329 | 331 |
| 0.10 | 331 | 322 | 334 | 333 | 327 | 327 | 327 | 331 | 3.57 | 0.38 | 0.97 | 329 | 331 |
| 0.20 | 328 | 319 | 330 | 330 | 322 | 324 | 326 | 328 | 3.70 | 0.39 | 1.05 | 326 | 328 |

The temperature dependences of the spontaneous magnetization $M_S(T)$ for determining the critical index $\beta$ in the $La_{0.8-x}\square_x Na_{0.2} Mn_{1+x} O_{3-\Delta}$ nanopowders with $x = 0.00$–$0.20$ are presented in Fig. S7.

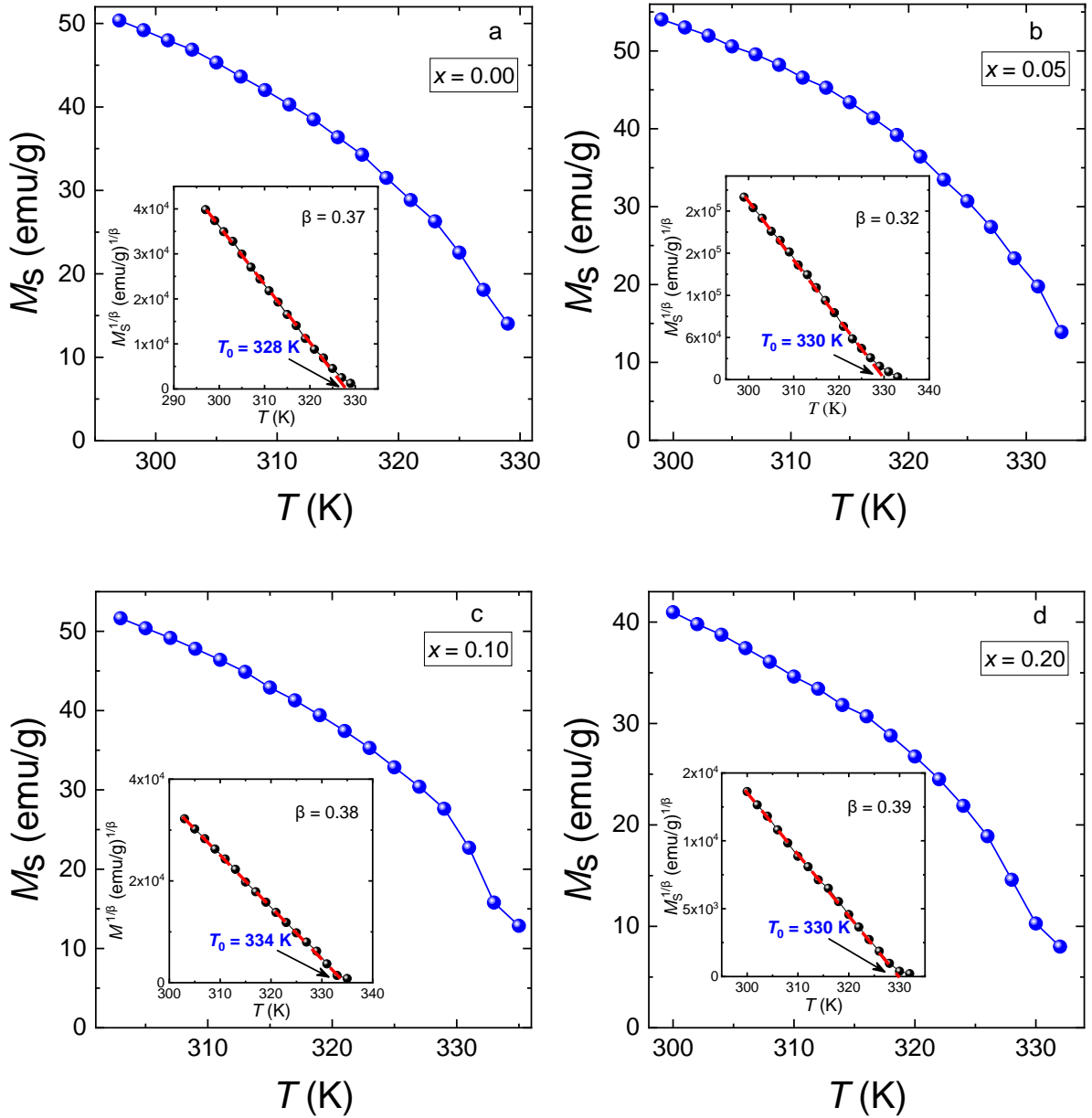

**Fig. S7.** Temperature dependences of the spontaneous magnetization $M_S(T)$ for the La$_{0.8-x}$□$_x$Na$_{0.2}$Mn$_{1+x}$O$_{3-\Delta}$ nanopowders with $x$ = 0.00 (a), 0.05 (b), 0.10 (c) and 0.20 (d). The inset shows the temperature dependence $M_S^{1/\beta}(T)$ with different critical indices of β and temperature $T_0$.

The Arrott plots $M^{1/\beta}((H/M)^{1/\gamma})$ with different models and temperature dependences of relative slope RS for the La$_{0.8-x}$□$_x$Na$_{0.2}$Mn$_{1+x}$O$_{3-\Delta}$ nanopowders with $x$ = 0.00–0.20 are presented in Fig. S8.

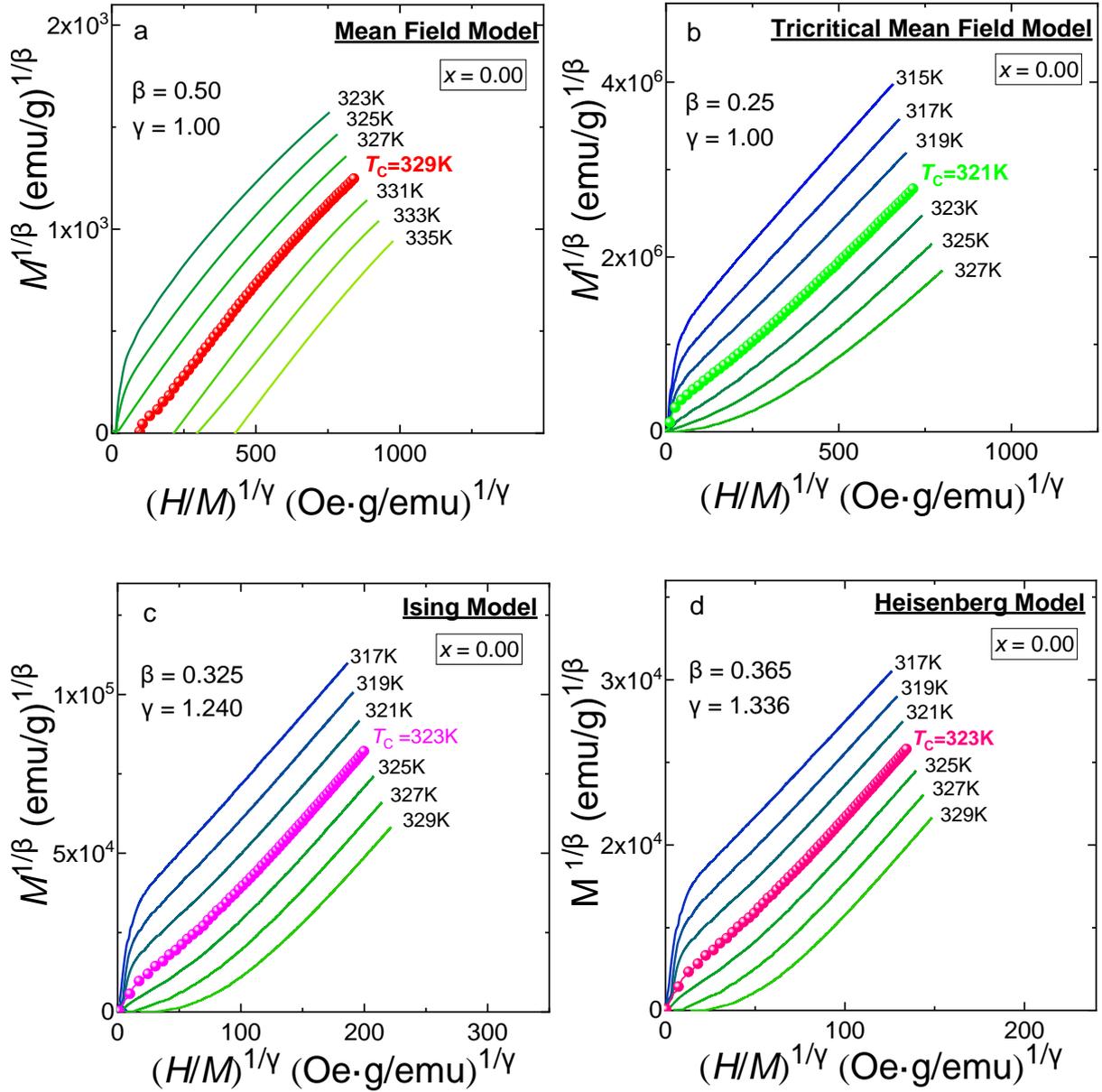

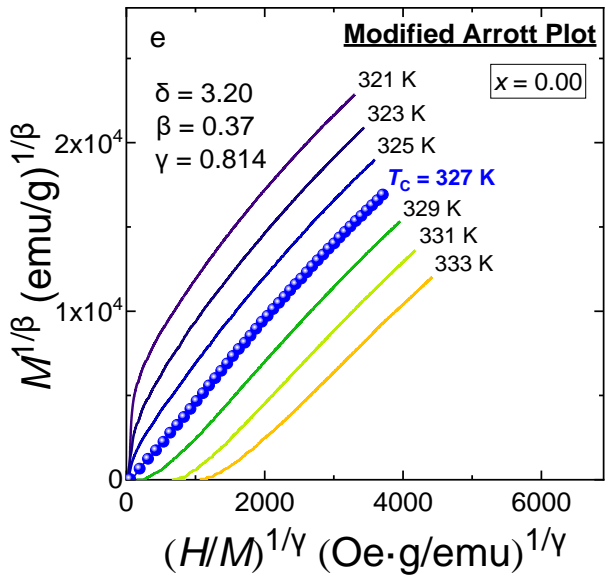
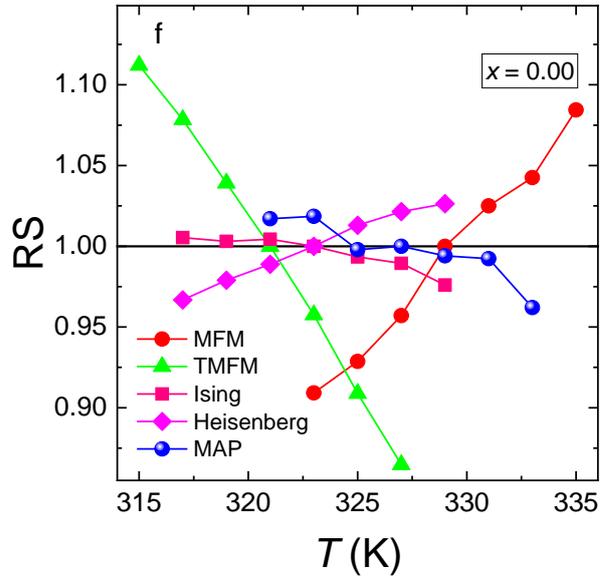
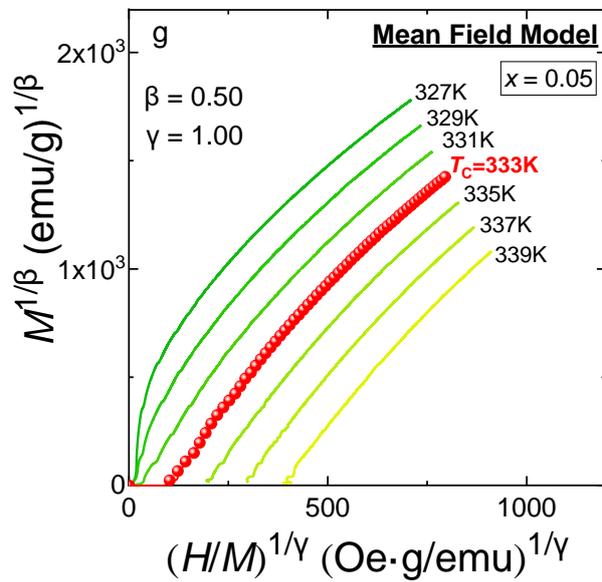
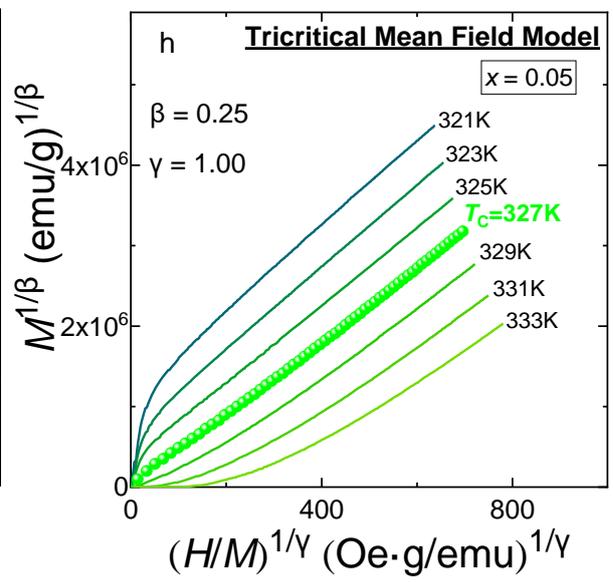
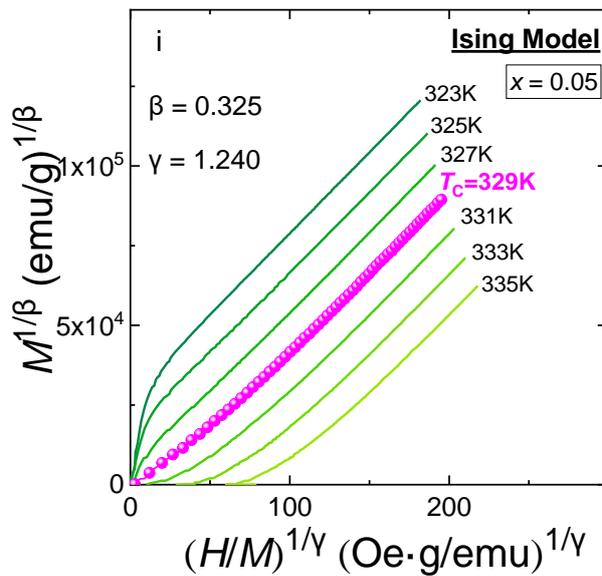
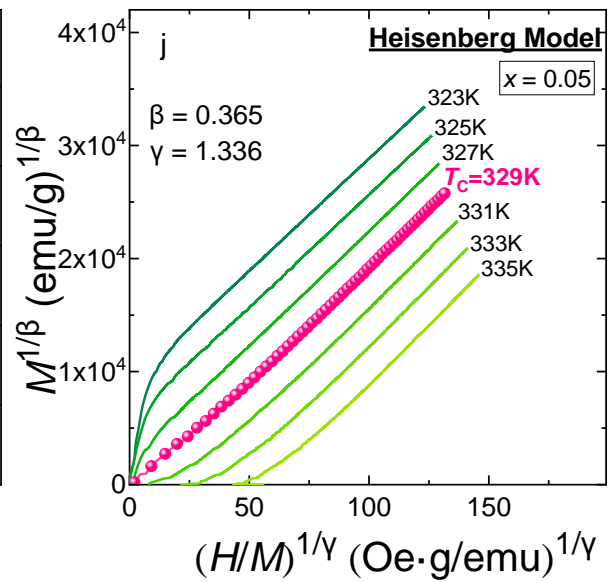

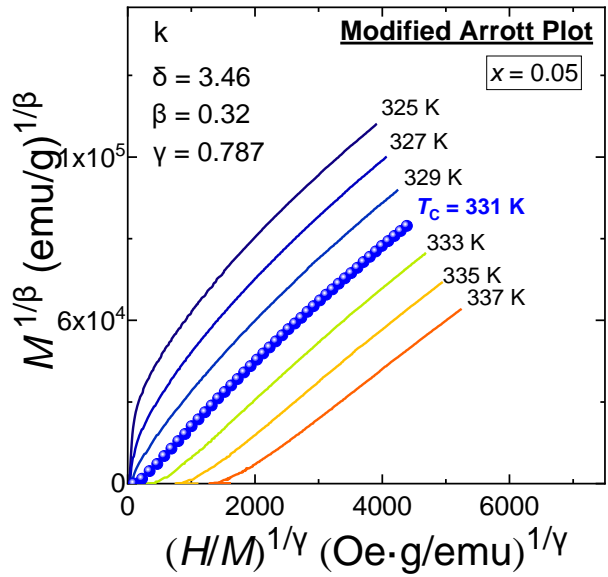
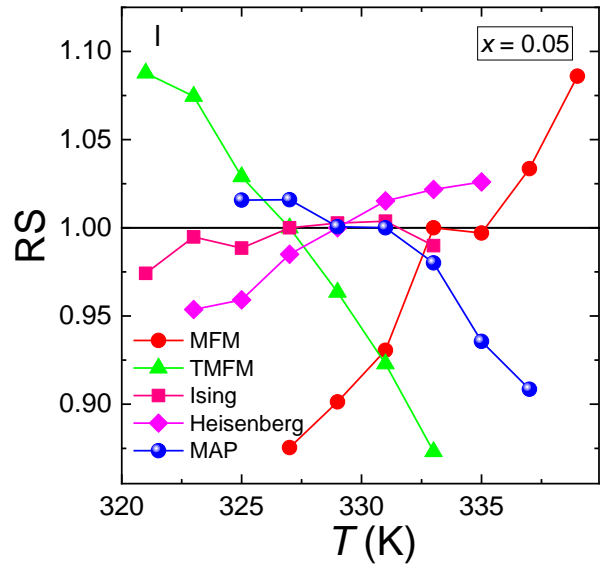
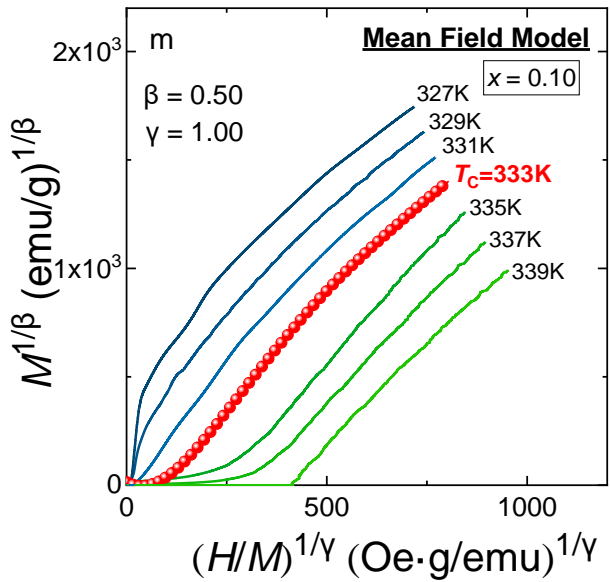
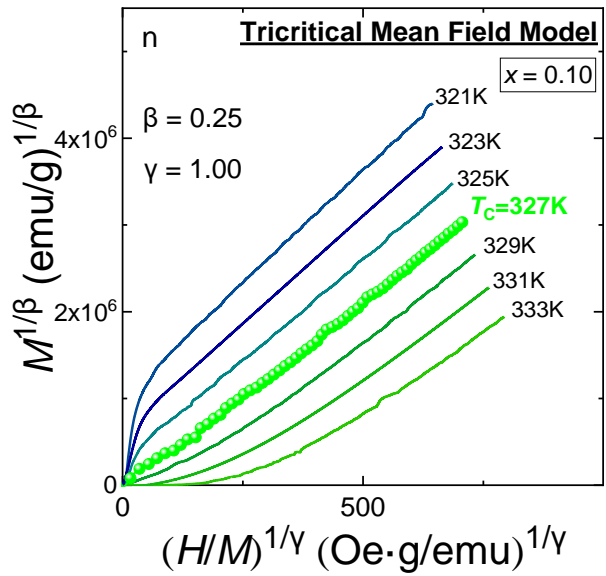
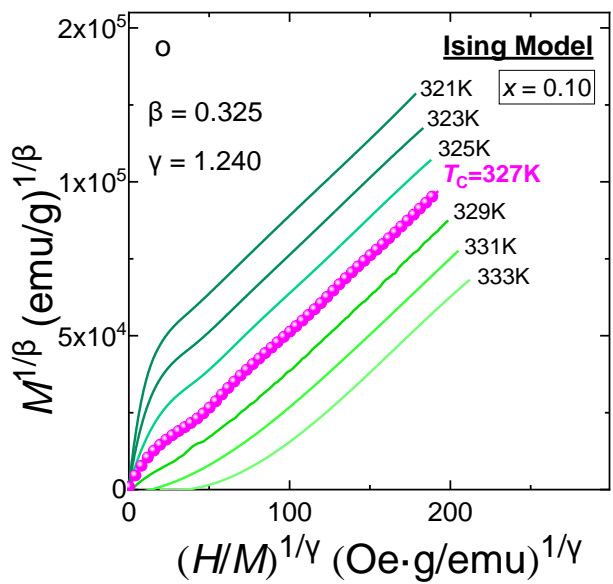
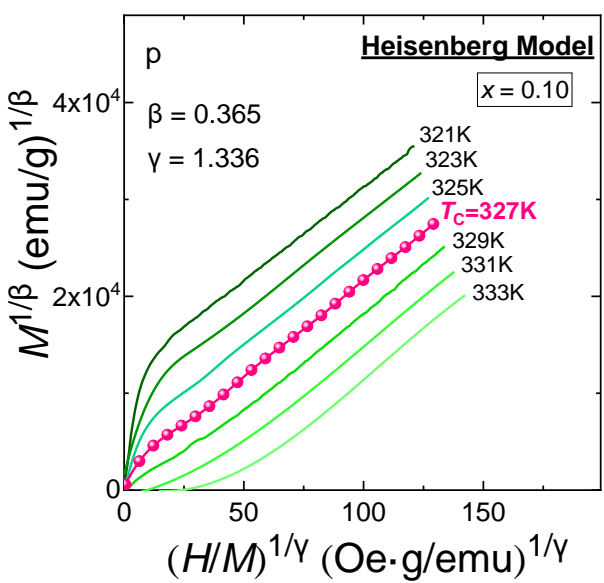

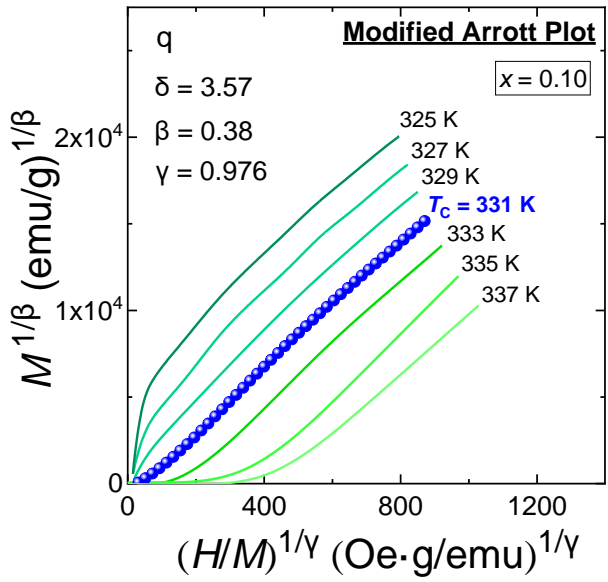
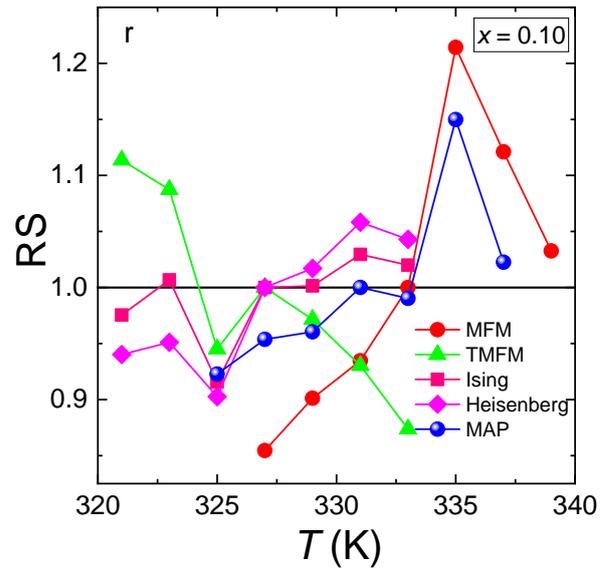
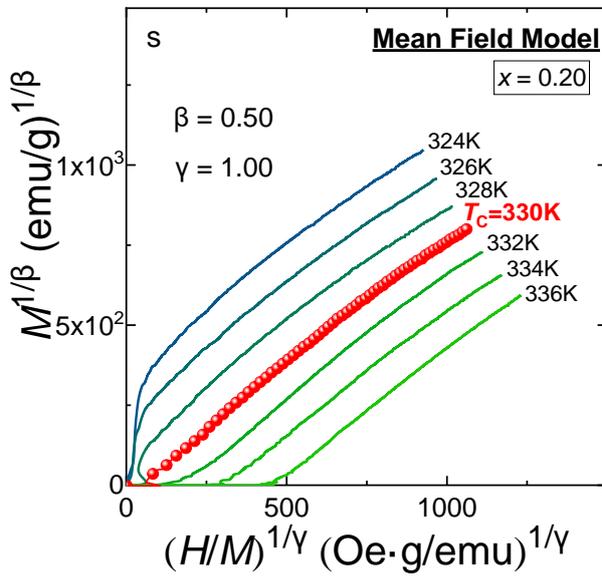
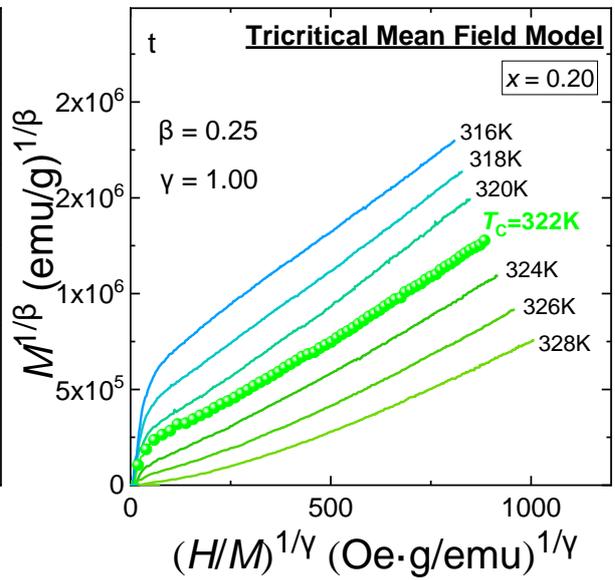
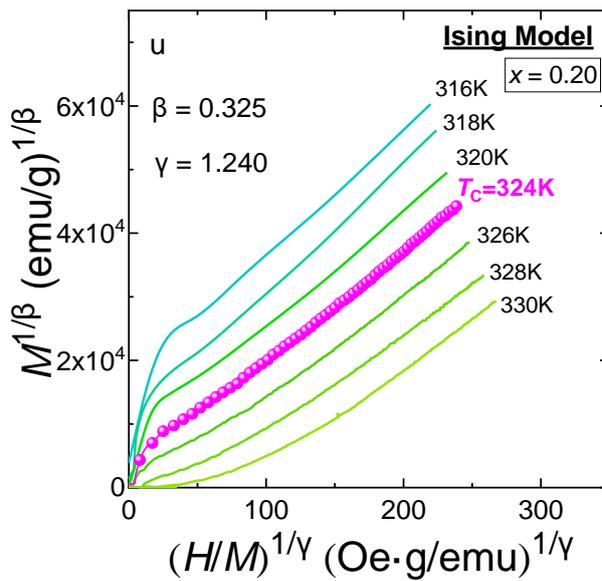
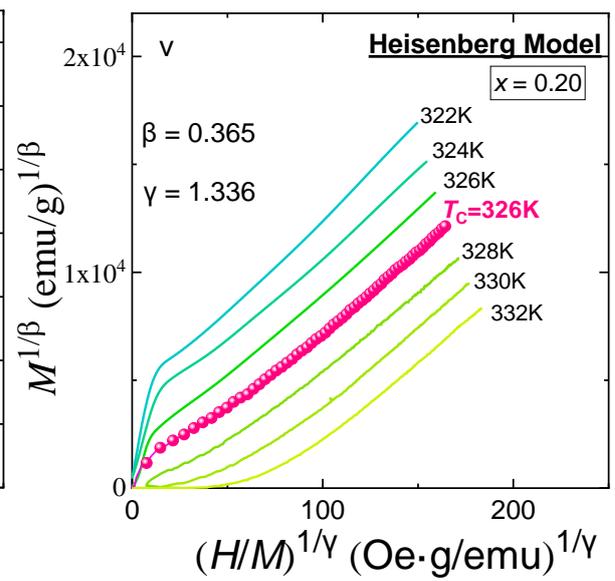

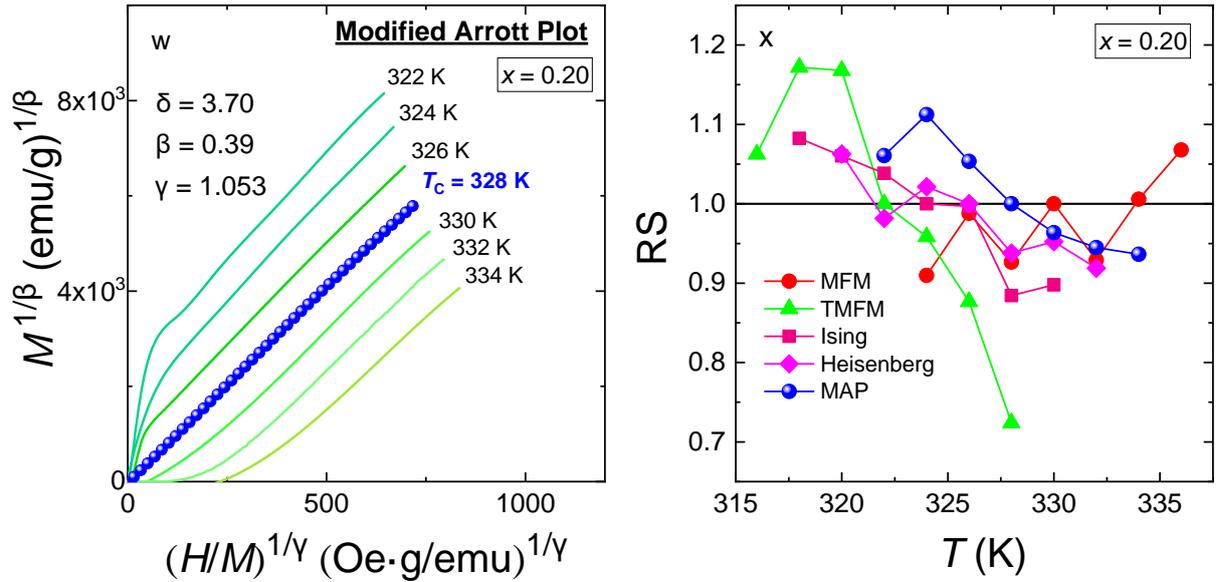

**Fig. S8.** Arrott plots $M^{1/\beta}((H/M)^{1/\gamma})$ with mean field model (a,g,m,s), tricritical mean field model (b,h,n,t), 3D-Ising model (c,i,o,u), 3D-Heisenberg model (d,j,p,v) and modified Arrott plot (e,k,q,w), as well as temperature dependence of relative slope (RS) (f,l,r,x) defined as RS = $S(T)/S(T_C)$ for the $La_{0.8-x}\square_x Na_{0.2}Mn_{1+x}O_{3-\Delta}$ nanopowders with $x$ = 0.00 (a-f), 0.05 (g-l), 0.10 (m-r) and 0.20 (s-x), respectively.

## S7. Magnetocaloric properties of the La$_{0.8-x}$□$_x$Na$_{0.2}$Mn$_{1+x}$O$_{3-\Delta}$ nanopowders

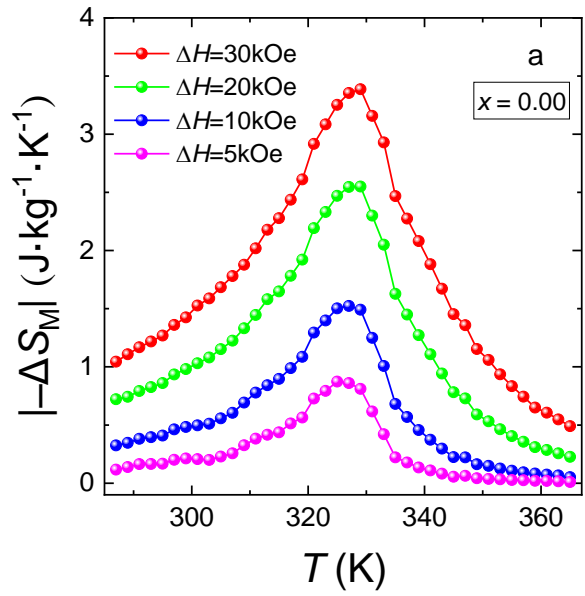
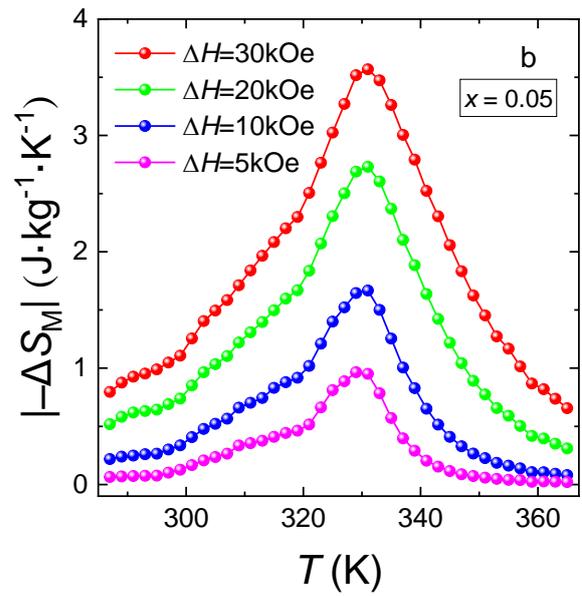
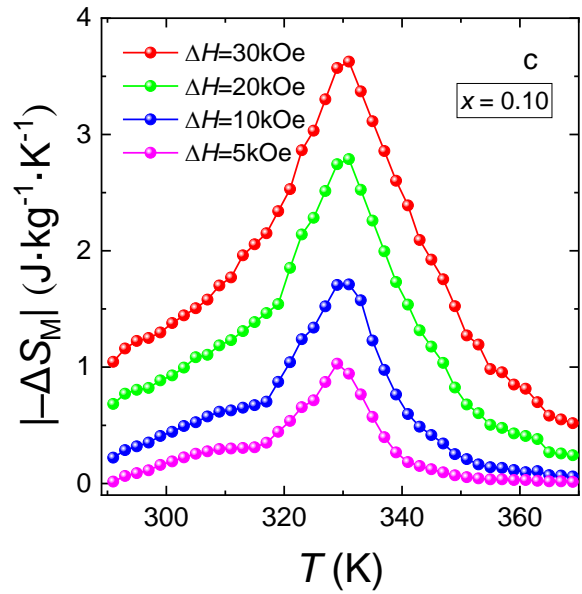
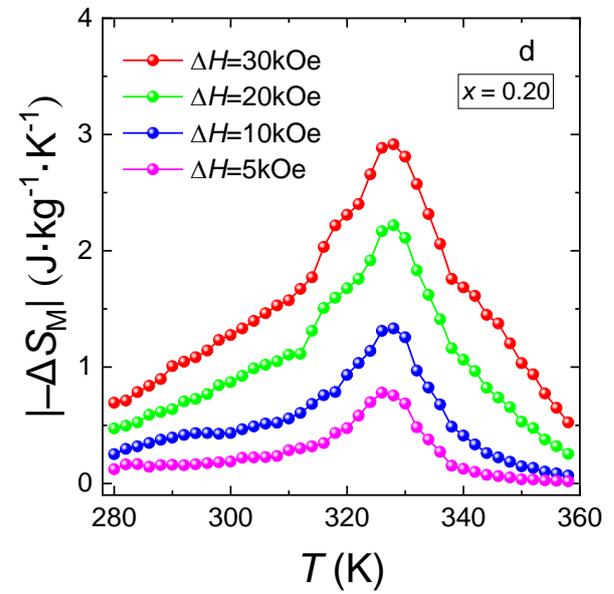

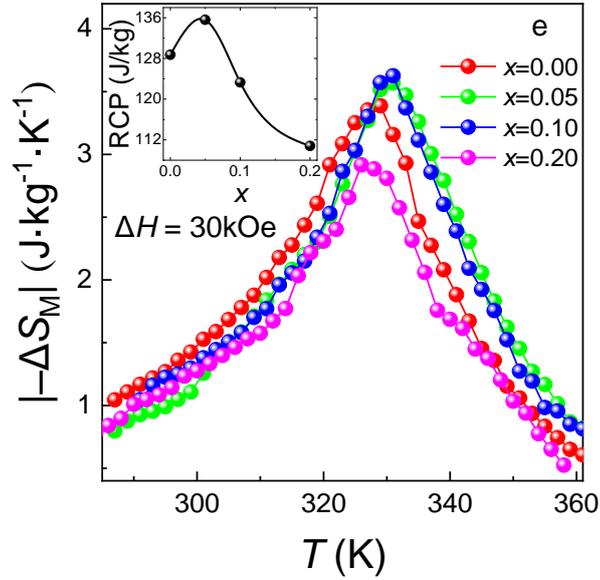

**Fig. S9**. Temperature dependences of the magnetic entropy change $-\Delta S_M(T)$ for the La$_{0.8-x}$□$_x$Na$_{0.2}$Mn$_{1+x}$O$_{3-\Delta}$ nanopowders with $x = 0.00$ (a), 0.05 (b), 0.10 (c) and 0.20 (d), as well as the comparable $-\Delta S_M(T)$ for all compositions (e) (the insert shows the concentration dependence of relative cooling power RCP($x$)).

For the simplest understanding contributions from SPM and FM particles in the MCE, Fig. S10 demonstrates temperature dependences of the normalized entropy $\Delta S/\Delta S_{max}(T)$ in the both small ($H = 2$ kOe) and strong ($H = 30$ kOe) magnetic field regions. The appearance of two MCE peaks at two different $T_{MCE}$ are observed and the difference between them is $\Delta T_{MCE} = 4$ K for $x = 0.00$, 2 K for $x = 0.05$, 2 K for $x = 0.10$ and 2 K for $x = 0.20$.

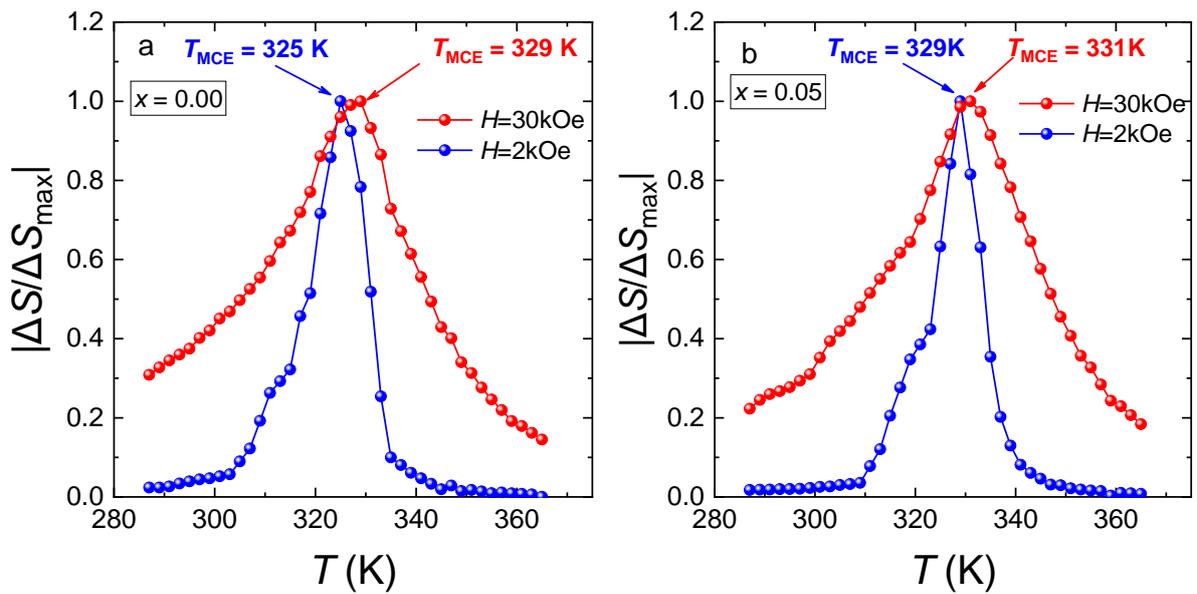

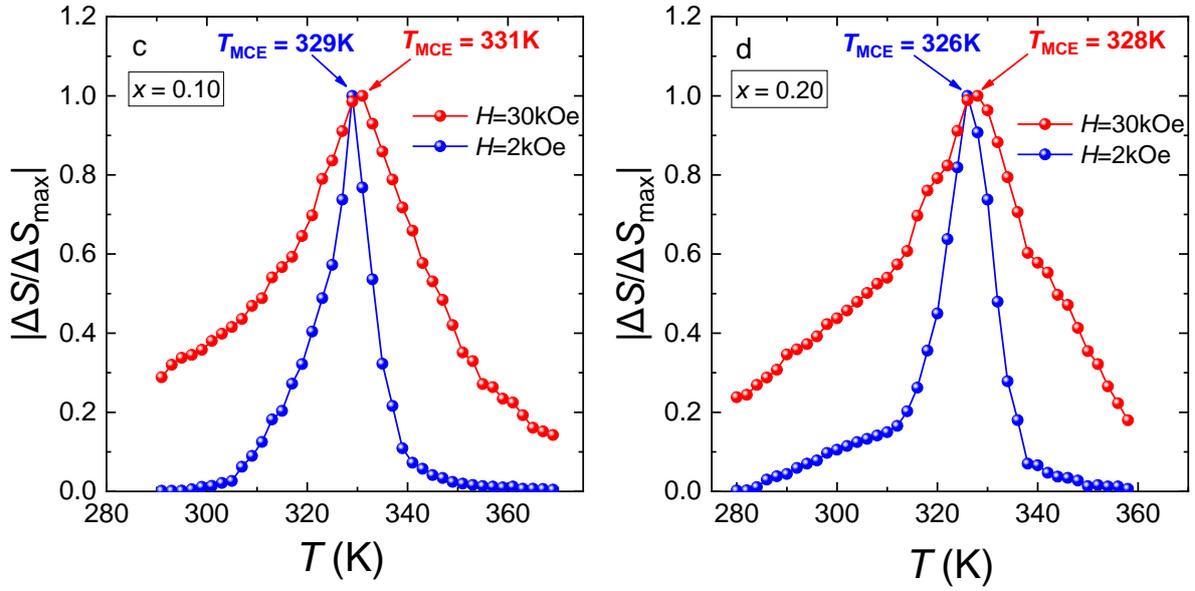

**Fig. S10**. Temperature dependences of the normalized entropy $\Delta S/\Delta S_{max}(T)$ for the $La_{0.8-x}\square_x Na_{0.2}Mn_{1+x}O_{3-\Delta}$ nanopowders with $x = 0.00$ (a), 0.05 (b), 0.10 (c) and 0.20 (d).

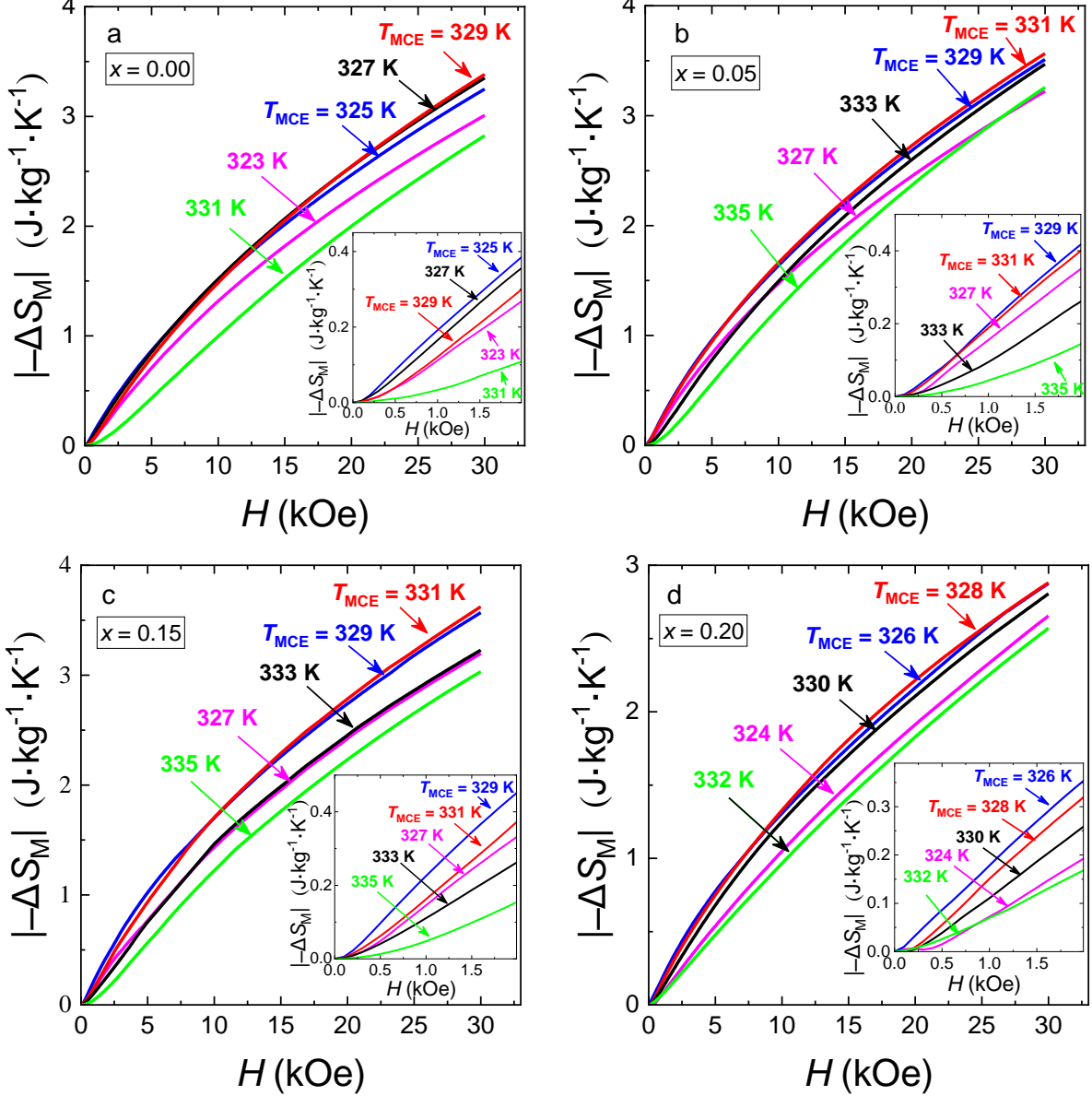

**Fig. S11**. Field dependences of the magnetic entropy change $-\Delta S_M(H)$ for the $La_{0.8-x}\square_x Na_{0.2}Mn_{1+x}O_{3-\Delta}$ nanopowders with $x = 0.00$ (a), 0.05 (b), 0.10 (c) and 0.20 (d). The inserts show field dependences $-\Delta S_M(H)$ in the small magnetic field region.

According to Table S12, the studied $La_{0.8-x}\square_x Na_{0.2}Mn_{1+x}O_{3-\Delta}$ compositions have comparable or even higher values of magnetocaloric parameters than other magnetic compositions (especially powders) that makes them promising ones for possible application in magnetic cooling system devices [29] and for treating cancer by local magnetic hyperthermia in medicine [30, 31]. The total data about MCE parameters for all studied samples are presented in Table S13. The highest value of MCE is noted for the $La_{0.7}\square_{0.1}Na_{0.2}Mn_{1.1}O_{3-\Delta}$ composition. At the same time, the relative cooling power (RCP) parameter $RCP = |-\Delta S_M^{max}|\cdot\Delta T_{FWHM}$, where $\Delta T_{FWHM}$ is the full-width at half maximum temperature, achieves the maximum value at $\mu_0\Delta H = 3$ T for the $x = 0.05$ (see insert in Fig. S9(e)). It should also be noted that a deviation from stoichiometry and the creation of cation and anion vacancies on *A*-, *B*- and oxygen-sublattices lead to an additional improvement in the magnetocaloric parameters (see Table S12). Moreover, overstoichiometric manganese and the creation of cation vacancies at approximately the same oxygen non-stoichiometry and particle size for the $La_{0.8-x}\square_x Na_{0.2}Mn_{1+x}O_{3-\Delta}$ nanopowders have a favorable influence on the MCE.

**Table S12**
The magnetic ordering temperature ($T_C$) and magnetocaloric parameters ($-\Delta S_M^{max}$, RCP) of the manganite $La_{0.8-x}\square_x Na_{0.2}Mn_{1+x}O_{3-\Delta}$ nanopowders in comparison with other magnetic materials

| Composition | $T_C$ (K) | $\mu_0\Delta H$ (T) | $-\Delta S_M^{max}$ (J/kg·K) | RCP (J/kg) | Size (nm) | Ref. |
|---|---|---|---|---|---|---|
| *(La-Na-Mn-O) system* | | | | | | |
| $La_{0.8}Na_{0.2}MnO_3$ | 297 | 1 | 1.25 | – | 33 | [32] |
| $La_{0.7}\square_{0.1}Na_{0.2}Mn_{1.1}O_{2.95}$ | 331 | 1 | 1.71 | 32 | 56 | This work |
| $La_{0.75}\square_{0.05}Na_{0.2}Mn_{1.05}O_{2.96}$ | 331 | 1 | 1.67 | 40 | 60 | This work |
| $La_{0.8}Na_{0.2}MnO_{2.96}$ | 327 | 1 | 1.52 | 35 | 61 | This work |
| $La_{0.8}Na_{0.2}MnO_3$ | 334 | 1 | 1.96 | 86 | bulk | [29] |
| $La_{0.799}Na_{0.199}MnO_{2.97}$ | 334 | 1 | 2.00 | 90 | bulk | [33] |
| *(La-Ag-Mn-O) system* | | | | | | |
| $La_{0.8}Ag_{0.2}MnO_3$ | 306 | 2 | 0.96 | 39.98 | 17 | [34] |
| $La_{0.6}Ag_{0.2}Mn_{1.2}O_3$ | 308 | 1 | 2.03 | 12 | 68 | [23] |
| $La_{0.8}Ag_{0.2}MnO_3$ | 300 | 1 | 2.40 | 32 | bulk | [29] |
| $La_{0.7}Ag_{0.2}Mn_{1.1}O_3$ | 271 | 1 | 2.46 | 32 | bulk | [35] |
| *(La-K-Mn-O) system* | | | | | | |
| $La_{0.85}K_{0.15}MnO_3$ | 274.5 | 2 | 2.02 | – | 40-50 | [36] |
| $La_{0.85}K_{0.15}MnO_3$ | 274.5 | 2 | 3.06 | – | 90-100 | [36] |
| $La_{0.85}K_{0.15}MnO_3$ | 274.5 | 2 | 3.56 | – | 120-130 | [36] |
| $La_{0.8}K_{0.2}MnO_3$ | 281 | 2 | 1.91 | 63 | powder | [37] |
| $La_{0.796}K_{0.196}Mn_{0.993}O_3$ | 344 | 1.5 | 2.19 | 84 | bulk | [38] |
| *(La-Sr-Mn-O) system* | | | | | | |
| $La_{0.67}Sr_{0.33}MnO_3$ | 275 | 5 | 1.092 | 133.13(3) | 48 | [39] |
| $La_{0.67}Sr_{0.33}MnO_3$ | 315 | 5 | 1.615 | 138.82(6) | 65 | [39] |
| $La_{0.67}Sr_{0.33}MnO_3$ | 350 | 5 | 1.791 | 229.16(8) | 85 | [39] |
| $La_{0.67}Sr_{0.33}MnO_3$ | 370 | 5 | 2.394 | 248.05(3) | 96 | [39] |
| $La_{0.67}Sr_{0.33}MnO_3$ | 348 | 5 | 1.69 | 211 | bulk | [40] |
| *(La-Ca-Mn-O) system* | | | | | | |
| $La_{0.6}Ca_{0.4}MnO_3$ | 258 | 5 | 2.3 | 228 | 45 | [41] |
| $La_{0.6}Ca_{0.4}MnO_3$ | 269 | 5 | 3.5 | 251 | 70 | [41] |
| $La_{0.6}Ca_{0.4}MnO_3$ | 272 | 5 | 5.8 | 374 | 122 | [41] |
| $La_{0.6}Ca_{0.4}MnO_3$ | 270 | 5 | 8.3 | 508 | 223 | [41] |
| $La_{0.6}Ca_{0.4}MnO_3$ | 263 | 3 | 5.0 | 135 | bulk | [42] |
| Gd | 293 | 1.5 | 4.20 | – | bulk | [38] |

**Table S13**
The magnetic entropy change ($-\Delta S_M^{max}$), the full-width at half maximum temperature ($\Delta T_{FWHM}$) and relative cooling power (RCP) for the $La_{0.8-x}\square_x Na_{0.2}Mn_{1+x}O_{3-\Delta}$ nanopowders with $x = 0.00–0.20$

| x | $-\Delta S_M^{max}$ (J/kg K) | | | | $\Delta T_{FWHM}$ (K) | | | | RCP (J/kg) | | | |
|---|---|---|---|---|---|---|---|---|---|---|---|---|
| | 0.5 T | 1 T | 2 T | 3 T | 0.5 T | 1 T | 2 T | 3 T | 0.5 T | 1 T | 2 T | 3 T |
| 0.00 | 0.87 | 1.52 | 2.55 | 3.39 | 18 | 23 | 31 | 38 | 16 | 35 | 79 | 128 |
| 0.05 | 0.96 | 1.67 | 2.73 | 3.57 | 16 | 24 | 32 | 38 | 15 | 40 | 87 | 135 |
| 0.10 | 1.03 | 1.71 | 2.79 | 3.63 | 16 | 19 | 27 | 34 | 16 | 32 | 75 | 123 |
| 0.20 | 0.78 | 1.33 | 2.22 | 2.92 | 17 | 22 | 30 | 38 | 13 | 29 | 67 | 110 |